%% file: main.tex
\documentclass[journal]{article}
\usepackage[margin=1 in]{geometry}

\usepackage[utf8]{inputenc}
\usepackage{etex}
\usepackage{graphicx}
\usepackage{fleqn}
\usepackage{cite}
\usepackage{siunitx}
\usepackage{amsmath,amssymb,amsfonts}
\usepackage{algorithmic}
\usepackage{graphics}
\usepackage{textcomp}
\usepackage{tikz}
\usepackage{color}
\usepackage{comment}
\usepackage{xcolor}
\usepackage{filecontents}
\usepackage{float}
\usetikzlibrary{arrows.meta,patterns}
\usepackage{graphics}
\usepackage{pict2e,color}
\usepackage{subcaption}
\usepackage{nccmath}
\usetikzlibrary{fit}
\usepackage{nccmath}
\usepackage{pgfplots}
\usepackage{lipsum}
\usepackage{stackengine}
\usepackage[capitalize,noabbrev]{cleveref}
\usepackage[ruled,vlined]{algorithm2e}
\usepackage{amsthm}
\usepackage{balance}
\usepackage{color,soul}
\usepackage{dsfont}
\newtheorem{theorem}{Theorem}

\setul{0.5ex}{0.3ex}
\setulcolor{blue}
\usetikzlibrary{positioning, arrows.meta,calc}
\def\BibTeX{{\rm B\kern-.05em{\sc i\kern-.025em b}\kern-.08em
    T\kern-.1667em\lower.7ex\hbox{E}\kern-.125emX}}

\def\H{\mathbf{H}}    
\def\J{\mathbf{J}}
\def\bGamma{\mathbf{\Gamma}}

\usepackage{mathtools}

\def\E{\mathbb{E}}

\def\ln{\mathrm{ln}}
\def\Fx{f(\pmb{x})}
\def\Cx{{c}(\pmb{x})}
\newcommand{\B}[1]{\mathbf{#1}} 

\usepackage{acro}
\DeclareAcronym{NN}{short=NN, long=neural network}
\DeclareAcronym{FFNN}{short=FFNN, long=feed-forward neural network}
\DeclareAcronym{GNN}{short=GNN, long=graph neural network}
\DeclareAcronym{GGNN}{short=GGNN, long=gated graph neural network}
\DeclareAcronym{GRU}{short=GRU, long=gated recurrent unit}
\DeclareAcronym{RL}{short=RL, long=reinforcement learning}

\usepackage[colorinlistoftodos,prependcaption,textsize=footnotesize]{todonotes}
\usepackage{regexpatch}
\makeatletter
\xpatchcmd{\@todo}{\setkeys{todonotes}{#1}}{\setkeys{todonotes}{inline,#1}}{}{}
\makeatother
\date{}
\begin{document}

\title{
Utility-driven Optimization of TTL Cache Hierarchies under Network Delays
}

 \author{
 {Karim S. Elsayed$^*$, Fabien Geyer$^\dagger$, Amr Rizk$^*$ } \\
\textit{$^*$Faculty of Computer Science, University of Duisburg-Essen, Germany} \\
\textit{$^\dagger$ Airbus Central R$\&$T, Munich, Germany}
}
\maketitle

\begin{abstract}
We optimize hierarchies of Time-to-Live (TTL) caches under random network delays. 
A TTL cache assigns individual eviction timers to cached objects that are usually refreshed upon a hit where upon a miss the object requires a random time to be fetched from a parent cache.
Due to their object decoupling property, TTL caches are of particular interest since the optimization of a per-object utility enables service differentiation. 
However, state-of-the-art exact TTL cache optimization does not extend beyond single TTL caches, especially under network delays.

In this paper, we leverage the object decoupling effect to formulate the non-linear utility maximization problem for TTL cache hierarchies in terms of the exact object hit probability under random network delays. 
We iteratively solve the utility maximization problem to find the optimal per-object TTLs.
Further, we show that the exact model suffers from tractability issues for large hierarchies and propose a machine learning approach to estimate the optimal TTL values for large systems.
Finally, we provide numerical and data center trace-based evaluations for both methods showing the significant offloading improvement due to TTL optimization considering the network delays.


\end{abstract}


\vspace{-5pt}
\section{Introduction}
\label{Sect:Intro}
%
Caching has an essential role in reducing response times and bandwidth consumption by drawing frequently requested data objects closer to the request origin~\cite{Ramesh_Akamai_hierarchy,Dist_alg_CDN}.
%
As the demand for objects dynamically varies, service differentiation becomes an integral building block of the caching system.
This is achieved through a caching utility abstraction~\cite{Deghan_utility} that gives rise to concepts such as caching fairness and cache resource allocation.
A key approach to fine-grained cache optimization lies in controlling the utility attained \textit{by individual  objects}.
This calls for systems built around \textit{independently} tuning object-specific cache performance metrics.

Traditional and modern caches that use either fixed or data-driven, learned policies~\cite{Traditional_caching_policies, Che:LRUtoTTL, DeepCache_popularity_prediction,FNN_popularity_prediction} usually run one or more decision rules that couple the occupancy of all the objects in the cache. 
For example, Least-recently-used (LRU) is a versatile and popular caching policy that is based on such coupling. 
In contrast, TTL caching decouples the object occupancy in the cache by assigning \textit{independent} expiry/eviction time tags to the individual objects~\cite{TTL_early_model}.
In a single TTL cache the aggregate utility of objects is maximized by deriving the object-specific optimal TTL values~\cite{Deghan_utility}. 
While the work in~\cite{Deghan_utility} is seminal, it is limited to a single cache within a set of restricting assumptions including, zero network delays and Poisson requests.


Caching systems usually consist of multiple caches connected to form a hierarchy~\cite{Ramesh_Akamai_hierarchy}. 
We consider tree-like hierarchies representing systems such as content delivery networks (CDN) and data center caching.
We note that optimizing cache hierarchies is hard as it involves \textit{jointly modeling and optimizing} multiple interconnected caches under random network delays.
It is also known that optimizing caches in isolation poorly utilizes the storage by allowing redundancy and ignoring the request process correlations~\cite{Cache_redund}.


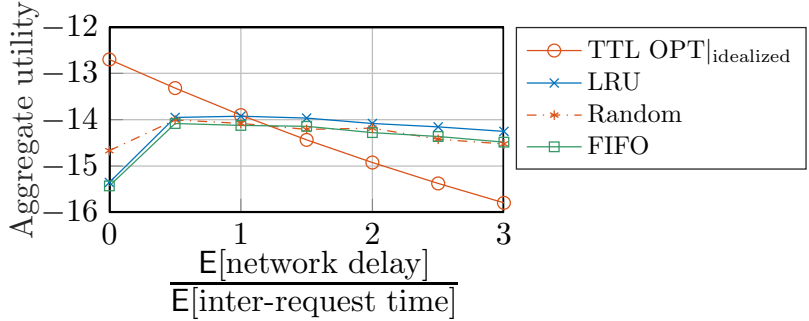
\begin{figure}[t]
         \centering
	 \resizebox{0.65\linewidth}{!}{\input{Evaluations/Delay_Impact_on_three_caches/Motivation}}
      \captionsetup{skip=0pt} 
     \caption[]{\textbf{Impact of delays on the TTL caching performance:} The aggregate utility\footnotemark of a 3-cache binary tree for increasing network delays and a fixed request rate at the leaves. The TTL caches are jointly optimized under the state-of-the-art zero network delay assumption. The performance of the TTL caching system deteriorates with network delays below delay-agnostic hierarchies built upon LRU, FIFO or random caching.}
     \label{fig:Obj_hitpr_delrat4_intro}
     \vspace{-15pt}
\end{figure}


Previous work on modeling and performance evaluation of TTL caching hierarchies offers computable models that are either exact-but-slow~\cite{Berger:TTL_MAP,Elsayed_cache_hierarchy_MAP} or approximate-but-fast~\cite{FofackNNT14,Dehghan_cache_with_delay}.
For single TTL caches there exist exact and computable models under the assumption of 
network delay randomness~\cite{Dehghan_cache_with_delay}.
Additionally, the work in~\cite{Don_hiera_ut} utilizes an approximate approach to simplify the TTL cache hierarchy and propose a corresponding analytical optimization. 
The work in~\cite{Berger:TTL_MAP} was the first to provide an exact Markov arrival process (MAP) model of a TTL caching tree under zero network latency.
\footnotetext{Utility captured through $\alpha$-fairness~\cite{Deghan_utility}, here, as $\sum_i \sum_j \lambda_{Iij} \log(P_{ij})$ with $\lambda_{Iij}$ and $P_{ij}$ being the average request rate and system hit probability of the Input request stream of object~$i$ arriving at leaf cache $j$, respectively.}
As empirically observed in~\cite{AtreSWB20}, for a wide set of caching systems, the network delay significantly impacts the caching performance. 
The authors observe that the network delay ranges from being in the order of the inter-request times to being multiple orders of magnitude larger. 
Fig.~\ref{fig:Obj_hitpr_delrat4_intro} shows the detrimental impact of ignoring the network delays when optimizing the exact analytical model (denoted ``idealized'').
Here, a binary tree of three TTL caches is optimized under the zero network latency assumption. 
It then performs significantly better than a size-comparable, delay-agnostic\footnote{\textit{Delay-agnostic} as these caching rules are fixed no matter the delay.} LRU, Random or FIFO cache hierarchy.
However, when this optimized TTL cache hierarchy is driven under random network delays,
its performance deteriorates significantly even below the performance of the delay-agnostic counterparts. 
Fig.~\ref{fig:Obj_hitpr_delrat4_intro} shows \textit{the need for considering network delays when optimizing TTL cache hierarchies.} 


In this paper, we optimize the aggregate utility of TTL caching hierarchies under random network delays.
With the trade-off between model complexity and accuracy in mind, we propose \textit{two approaches} for utility maximization.
First, we propose an \textit{analytical closed-form} approach based on the exact MAP model where we solve the corresponding non-linear utility optimization problem using interior-point optimization.
Second, we propose a \textit{\ac{GNN}} approach that models the caching hierarchy as a graph to predict the optimal object TTLs after being iteratively learned using reinforcement learning.
The rationale behind showing two approaches lies in the observation that 
with increasing number of objects $N$ and the number of caches $n_c$,  the analytical closed-form solution becomes intractable. 
Hence, we empirically show this \textit{transition} from the exact closed-form solution to the approximate GNN-based solution with increasing system size. 
Our contributions are:
\begin{itemize}
    \item We formulate a utility maximization problem for optimizing  TTL caching hierarchies under random network delays to uniquely \textit{leverage} the object decoupling effect of TTL caches. We provide an analytical solution to the non-linear utility maximization  (Sect.~\ref{Sect:system model}-\ref{Sect:Interior point}).
    \item As analytical TTL models only use  the expected cache storage utilization we propose a corresponding TTL policy that reduces the loss in the aggregate utility.
    We show numerically in Sect.~\ref{Sect:evaluations} that the simulated aggregate utility is close to the analytically obtained one.
    \item We provide a \ac{GNN} model to maximize the utility of the caching system for large hierarchies (Sect.~\ref{Sect:GNN}).
    \item We provide numerical evaluations of the analytical and the \ac{GNN} utility maximization methods (Sect.~\ref{Sect:evaluations}). 
\end{itemize}
Before we delve into the contributions, we first give an overview of the related work on modeling TTL cache hierarchies as well as cache utility maximization in the next section. 

\section{Background \& Related Work}
\label{Sect:background}

\subsection{Approaches to TTL cache modelling and optimization}

Utility functions that are widely used in the context of network resource allocation \cite{Kelly_Congestion, Srikant_control} have been recently adopted to caching models as a means to achieve service differentiation~\cite{Deghan_utility}. 
As the TTL policy decouples objects in the cache it is particularly well suited for utility maximization problems that control individual object performance metrics such as object hit probability and occupancy.
Note that capacity-driven caching policies such as LRU or FIFO can be emulated using TTLs by linking the capacity to the expected cache occupancy \cite{Che:LRUtoTTL,Berger:TTL_MAP,Fofack:TTL,JiangNT18,Gelenbe73a}.
The authors of \cite{NegliaCM18} study linear utility maximization for different caching policies, while the authors in~\cite{TTL_heavy_tailet_opt} consider the TTL utility maximization under heavy-tailed request processes and the works~\cite{Towsley_rate_vs_prob,Towsley_rate_vs_prob_Hazard} consider utilizing hit rates instead of probabilities for utility maximization.

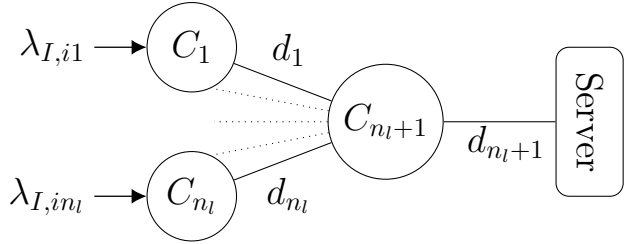
\begin{figure}[t]
\centering
\rotatebox{-90}{%
  \resizebox{0.2\linewidth}{!}{\input{Figs/tree}}
  }
        \captionsetup{skip=0pt} 
\caption{Two level caching hierarchy tree consisting of $n_l$ leaves. The random variable $d_i$ encodes the  delay for cache $i$ to download the content from its parent. The server contains all objects. The request rate of object $i$ at leaf $j$ is denoted $\lambda_{I,ij}$. }
\label{fig:caching_tree}
\end{figure}

The authors of \cite{AtreSWB20} draw attention to the impact of object retrieval delays in a \textit{single cache} and develop an algorithm similar to the clairvoyant Belady-Algorithm that provides an upper bound to the hit probability under network delays.
An exact analytical model for a single TTL cache under network delays is derived in \cite{Dehghan_cache_with_delay}.
For hierarchies, the authors of~\cite{Caching_hierarchy:Web_caching, Caching_hierarchy:CRAN_cooperative_caching,Caching_hierarchy:CDN,Caching_hierarchy:IPTV_video_cahcing ,Caching_hierarchy:IPTV} focus on optimizing the aggregate sum of the object utility of each cache  in terms of content placement and request routing.
Using variants of the fixed strategy Move-Copy-Down (MCD), the authors of~\cite{Don_hiera_ut} consider Poisson requests and use an approximation of the system hit probability, that, however, cannot capture the aggregation of different request streams, to optimize the overall utility of the hierarchy. 
Early works on TTL caching hierarchies use approximations to derive  performance metrics, e.g., the authors of \cite{Fofack:TTL} model the TTL caching hierarchies  assuming a renewal request process at the output of each cache.
An exact model of TTL cache hierarchies has been first proposed in~\cite{Berger:TTL_MAP} using a Markov Arrival Process (MAP) model under the zero network delay assumption.
Recently, the work in \cite{Elsayed_cache_hierarchy_MAP} extends this model to random network delays within the hierarchy.
Fig.~\ref{fig:caching_tree} depicts a two level cache hierarchy tree showing the object request processes at the leaves and the random network delays. 


In contrast to exact analytical models, learning-based approaches have been proposed for cache optimizations due to their lower computational complexity and ability to adapt to temporal changes.
The authors in~\cite{DeepCache_popularity_prediction} use a recurrent neural network with long short-term memory units to predict the future popularities of the objects and optimize the single cache hit probability.
In \cite{RL_cache,Zhong2018deep,Sadeghi2017optimal}, reinforcement learning (RL) approaches using feedforward neural networks are proposed for making content admission decisions for an individual cache. 
Recently, the work in \cite{Pedersen2021dynamic} proposed a deep RL (DRL)-based caching policy for distributed coded soft-TTL caching scenario with the goal to minimize the network load. 
In addition, the authors of~\cite{Utility_DRL} proposed a DRL online caching approach as an alternative to the analytical approach in~\cite{Deghan_utility} for a single utility-driven TTL cache to achieve faster adaptation to changes of content popularities.
In a TTL caching system that allows elastic adjustment of the cache storage size, a DRL-based algorithm can be used to optimize the system utility and the cost of storage together as outlined in~\cite{Utility_DRL_elastic_caching}.

Our work here differs fundamentally from the utility-driven caching in~\cite{NegliaCM18,Deghan_utility,TTL_heavy_tailet_opt,Towsley_rate_vs_prob,Utility_DRL,Towsley_rate_vs_prob_Hazard} as we consider the joint utility maximization for multiple caches in a hierarchy under random network delays.
In contrast to these related works, we base our work on an \textit{exact model of TTL cache hierarchies under random network delays}.
Further, in comparison, the work in~\cite{Don_hiera_ut} maximizes the system utility of a simplified hierarchy model with decoupled paths and approximate system hit probability.
In contrast, our work, first, takes random network delays into the optimization, and second, we build on the exact object hit probability throughout the hierarchy. Hence, instead of a fixed object placement such as MCD in~\cite{Don_hiera_ut} \textit{our optimization yields the optimal object TTLs at every cache in the hierarchy}. 
%
Finally, the main difference to the hit-driven optimizations for a single cache~\cite{DeepCache_popularity_prediction,FNN_popularity_prediction,RL_cache} is that our main goal is to provide utility maximization for entire hierarchies.






\begin{figure}[t]
\centering
        \hspace{-1cm}
      \begin{subfigure}[b]{0.4\linewidth}
      \vspace{-0.5cm}
      \resizebox{0.6\linewidth}{!}{\input{Figs/MAP_single_1 }}
      \centering
       \caption{}
        \label{Single_cache_MAP_ex1}
     \end{subfigure}
      \hspace{0.2cm}
      \begin{subfigure}[b]{0.4\linewidth}
            \centering
         \resizebox{0.9\linewidth}{!}{\input{Figs/MAP_Single_2.tex}}
        \caption{}
        \label{Single_cache_MAP_ex2}
     \end{subfigure} 
\caption{The MAP of a single cache object with parameters $\lambda$, $\mu$ and $\mu_F$ for the inter-request time, TTL and delay distributions, respectively. 
States "0", "1" and "F" indicate that the object is in the cache, the object is out of the cache and the object is being fetched, respectively.
The TTL and the delay are exponentially distributed. (a) represents a Poisson request process, while (b) represents an Erlang-2 request process. State pairs $X_a$ and $X_b$ correspond to the object during each of the two Erlang-2 request phases.}
    \label{Single_cache_MAP_ex}
\end{figure}
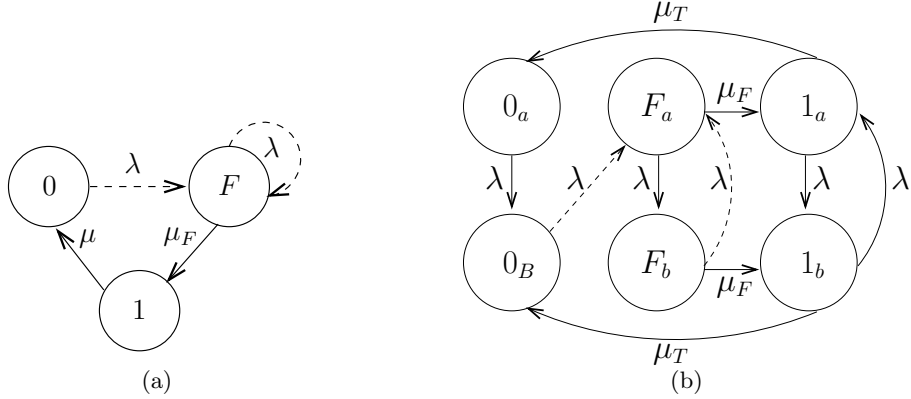

\subsection{Primer on exact TTL cache hierarchy modelling}
\label{sec:primer_TTL_exact}
Next, we provide a quick review of the exact analysis of TTL caching hierarchies given random network link delays. A detailed treatment is found in~\cite{Elsayed_cache_hierarchy_MAP}. 
We use bold symbols to denote vectors and matrices and non-bold symbols for scalars.  
Since TTL caches decouple objects, the model expresses the \textit{object-specific hit probability} using a MAP that describes the state of an object of interest in the hierarchy. 
A MAP~\cite{Asmussen} is a point process described by two transition matrices; a hidden one $\mathbf{D}_{{0}}$, and an active one $\mathbf{D}_{{1}}$. 
Let the states associated with these matrices be denoted by the vector $\pmb{S}$.
Transitions that only control the jump process $J(t)$ are in  $\mathbf{D}_{{0}}$, while $\mathbf{D}_{{1}}$ contains transitions which control $J(t)$ and also a counting process $N(t)$.
Here, $N(t)$ counts the request misses from the entire cache hierarchy (so called system misses) until time $t$.
Fig.~\ref{Single_cache_MAP_ex} depicts a state space model showing the state of the object in a single cache.
The figure shows different examples of the MAP of a single cache with Poisson and Erlang request processes while for simplicity TTLs and delays are exponentially distributed.



In general, the state space model is given as a composed MAP that is constructed of the MAPs of the request process, the TTL and delays, i.e., each of these three components can itself be described by a Markov arrival  process~\cite{Elsayed_cache_hierarchy_MAP}.
The MAP describing the state of object $i$ is, thus, defined over the state vector $\pmb{S_i}$ consisting of $n_{s,i}$ states.
To obtain the MAP associated with the object state in a tree-like cache hierarchy, the approach in~\cite{Elsayed_cache_hierarchy_MAP} iteratively groups the individual cache MAPs using Kronecker superposition operations.
The operation builds an equivalent MAP for multiple cache MAPs under topological constraints.

The steady state hit probability of an object $i$ given the caching hierarchy MAP is expressed as~\cite{Berger:2014}
\vspace{-2pt}
\begin{equation}
   P_i=1-\frac{\lambda_{Mi}}{\sum_{j=1}^{n_l} \lambda_{Iij}}
   =
   1-\frac{\pmb{\pi}_i  \mathbf{D}_{i,1} \pmb{1} }{  \sum_{j=1}^{n_l} \pmb{\pi}_{ij}^{\{I\}}  \mathbf{D}_{ij,1}^{\{I\}} \pmb{1}} \, ,
   \label{eq:hitpr_MAP}
\vspace{4pt}
\end{equation}
where $\lambda_{Mi}$ and $\lambda_{Iij}$ are the miss rate from the hierarchy, i.e. at the root, and the input request rate to the $j$-th leaf, respectively. 
The number of leaves is denoted by $n_l$.
Here, $\mathbf{D}_{{i,1}}$  and $\pmb{\pi}_i$  are the active transition matrix and the steady state probability vector of the hierarchy MAP, respectively. 
Further, $\mathbf{D}_{{ij,1}}^{\{I\}}$ and $\pmb{\pi}_{{ij}}^{\{I\}}$ are the active transition matrix and the steady state probability vector of the input request process at the $j$th leaf cache. 
The vector $\pmb{1}$ is an all-one column vector.

The steady state probability vector $\pmb{\pi}_i$ is associated with the MAP describing the state of object $i$ in the hierarchy.
Assuming stationary request processes, TTL sequences and delay processes, $\pmb{\pi}_i$ is classically found as
\vspace{-3pt}
\begin{equation}
\label{eq:steady_state_calc}
    \pmb{\pi}_i=\pmb{b} \pmb{A}_i^{-1}\;,
\vspace{4pt}
\end{equation}
where $b=[ 0,0,....,1]$ and $\pmb{A}_i=[(\mathbf{D}_{{i,0}}+\mathbf{D}_{{i,1}})_{:,n_{s,i}-1} |\pmb{1}]$.
Here, $(.)_{:,n_{s,i}-1}$ represent the first $n_{s,i}-1$ columns, i.e, all the columns except for the last one. The operation $[ \pmb{X} |\pmb{1}]$ concatenates the matrix $\pmb{X}$ and the column vector $\pmb{1}$.

The expected object occupancy $\E[O_i^k]$ of object $i$ in cache~$k$ in the hierarchy is
\vspace{-10pt}
\begin{equation}
    \E[O_i^k]=\sum_{j \in \chi_i^k} \pi_{i,j} \;,  
    \label{eq:Occ_MAP}
\end{equation}
where $\pi_{i,j}$ is the $j$th element of $\pmb{\pi}_i$.
Here, $\chi_i^k$ is the set of indices of states in $\pmb{S}_i$ where the object is stored in cache $k$.

Finally, note that it is evident from~\eqref{eq:steady_state_calc} and as highlighted in~\cite{Elsayed_cache_hierarchy_MAP} that obtaining $\pmb{\pi}_i$ is computationally intensive. 
With an increasing number of objects $N$ and number of caches $n_c$ the closed-form solution tractability is constrained by the matrix inversion $\pmb{A}_i^{-1}$.
Adopting the results from~\cite{Elsayed_cache_hierarchy_MAP} a direct computation shows that this solution scales as $\mathcal{O}(N \cdot N_s^{\xi n_c})$ with $\xi$ being the typical exponent for matrix inversion (e.g. from~\cite{CW_MatInv_complexity}) and $N_s$ being the number of states of each individual cache MAP assuming homogeneity.

\section{System model and Problem statement}
\label{Sect:system model}
In this section, we formulate the utility maximization problem for TTL cache hierarchies under random delays and discuss different notions of utility functions.
We consider a TTL caching tree with $n_c$ caches of which $n_l$ are leaf caches and the number of equally-sized objects in the system is $N$.
Exogenous object requests are only received via leaf nodes.
Object retrieval from one cache to a parent cache takes a non-trivial, random download delay.
We model each of the request processes, TTLs and the delays using a PH distribution.
We define a request to a leaf cache as a \textit{hit} when the object is in any of the caches along the path from that leaf to the root and as a miss otherwise. 
We use the approach outlined in the previous section to derive the MAP of the hierarchy.

\subsection{Key insight and approach to TTL optimization}
Next, we introduce our key idea for formulating the object utility given a cache hierarchy.
We aim at maximizing the aggregate utility of objects in the cache hierarchy.
This is achieved by controlling each object's overall \textit{hierarchy hit probability} through tuning the corresponding object \textit{TTL parameters at every cache} in the hierarchy.
%

In a cache hierarchy, requests to the same object arrive at heterogeneous leaf caches, where heterogeneity is in terms of leaf-specific object request processes as well as the object fetch delay distributions between child-parent caches.
Hence, requests of an object at different leaf caches contribute to the overall object utility differently.
The key idea of our formulation is to\textit{ decompose the object utility into its per input stream weighted utilities} and express it as
\begin{equation}
    U_i=\sum_j \lambda_{Iij} \psi(P_{ij}(\pmb{\mu}_i))\;,
    \label{eq: utility_decomp}
    \vspace{6pt}
\end{equation}
where $P_{ij}$ is the hierarchy hit probability of an object $i$ considering only the hits collected from the request stream to a leaf cache $j$, and the vector $\pmb{\mu}_i$ denotes the parameters of the given TTL distributions
of object $i$ at all caches in the hierarchy. 
Here, $\psi$ is a real-valued utility function, e.g. expressing $\alpha$-fairness as considered later on, and $\lambda_{Iij}$ is the request rate for object~$i$ at leaf cache~$j$. 
We calculate $P_{ij}$ from the MAP of object $i$ as
 \begin{equation}
   P_{ij}=1-\frac{\lambda_{Mij}}{ \lambda_{Iij}}
   =
   1-\frac{\pmb{\pi}_i  \mathbf{D}_{{ij,1}} \pmb{1} }{  \pmb{\pi}_{{ij}}^{\{I\}}  \mathbf{D}_{{ij,1}}^{\{I\}} \pmb{1}} \, ,
   \label{eq:hitpr_MAP_perleaf}
   \end{equation}
where only the miss rate $\lambda_{Mij}$ of the requests to leaf cache~$j$ is considered\footnote{Note from the formulation that the hierarchy hit probabilities per-leaf request stream are \emph{not mutually independent}.}. 
This is calculated using the active transition matrix $\mathbf{D}_{{ij,1}}$ containing only the transitions from  $\mathbf{D}_{{i,1}}$ that generate misses due to a request to leaf cache~$j$.
Note that $P_{ij}$ does not only depend on the object TTL parameters at the caches along the request stream path to the root but also on the object TTL parameters at each cache of the hierarchy.
This is expressed in \eqref{eq:hitpr_MAP_perleaf} by $\pmb{\pi}_i$ which is a function of all the parameters of the MAP.
Our formulation of the utility function in~\eqref{eq: utility_decomp} coincides with the utility definition in~\cite{Deghan_utility} in case of a single cache.
However, our formulation extends the utility calculation to tree-like cache hierarchies.

\subsection{Utility maximization}
\label{sect: utility_max}
We formulate the optimization of the object TTLs within the cache hierarchy given random network delays as a utility maximization problem. 
We express the overall utility of the hierarchy as the sum of the object utilities given in \eqref{eq: utility_decomp}.
We utilize the per-request stream decomposition formulation and the expected object occupancy in a specific cache in the hierarchy from~\eqref{eq:Occ_MAP}-\eqref{eq:hitpr_MAP_perleaf}.
Next, we formulate the following optimization problem
\begin{align}
    \underset{\pmb{\pmb{\mu}_i}}{\text{maximize}} \ & \sum_i^N \sum_j^{n_l} \lambda_{Iij} \psi(P_{ij}(\pmb{\mu}_i)) \;,
    \label{eq:obj_fn_final}\\[-0.25em]
\text{subject to} &\sum_i^N \E[O^k_{i}] =B_k \;  \  \forall k \;.
\label{eq:Main_opt_prob_avg_occupancy_constraint}
\end{align}
Here,  $\E[O^k_{i}]$ is the expected  occupancy of object~$i$ at cache~$k$ from~\eqref{eq:Occ_MAP} and $B_k$ denotes the size of cache $k$ measured in the number of objects the cache can hold.


Standard TTL models such as~\cite{Che:LRUtoTTL,Fofack:TTL, Berger:TTL_MAP, Deghan_utility} assume an infinite storage capacity and use the constraint~\eqref{eq:Main_opt_prob_avg_occupancy_constraint} on the expected occupancy to introduce a separate cache capacity constraint. 
In this basic formulation of the optimization problem, we take the same approach. 
We note that this constraint is not sufficient to prevent a storage capacity overflow or in other words a mismatch between the closed-form solution of~\eqref{eq:obj_fn_final} and an algorithmic realization using any finite cache size.
In Sect.~\ref{Sect:meTTL} we illustrate the problem of cache storage violation and propose an approach based on the chosen TTL policy.

As discussed in Sect.~\ref{Sect:Intro}, TTL caching has the advantage of tuning the hit probability of each object independently. 
Thus, we aim to find the object TTLs at each cache that maximize the overall utility of the cache hierarchy.



\subsection{Utility and fairness in cache hierarchies}
\label{sec:fairness}

Here, we adopt the $\alpha$-fair utility function proposed in~\cite{Walrand:Alpha_fair}, which groups different notions of fairness for resource allocation.
We use the $\alpha$-fair utility function
\begin{equation} \label{eq:alphafair_utility}
    \psi(P)=
    \begin{cases}
        \frac{P^{1- \alpha} }{1-\alpha}  \; \alpha \geq 0, \ \alpha \neq 1
        \\
        \log P, \; \alpha=1
    \end{cases}
    \vspace{6pt}
\end{equation}
where $P$ is the hit probability. 
The $\alpha$-fair utility function is used in the context of resource allocation to unify different notions of fairness~\cite{Srikant_control}. 
By letting $\alpha \in \left\{0,1\right\}$ or $\alpha \rightarrow\infty$, the $\alpha$-fair utility can represent the notion of fairness ranging from throughout (also called offloading) maximization (no fairness, $\alpha=0$) and proportional fairness ($\alpha=1$) to max-min fairness as $\alpha \rightarrow \infty$.
By solving the optimization function~\eqref{eq:obj_fn_final},~\eqref{eq:Main_opt_prob_avg_occupancy_constraint} we obtain the TTL values that maximize the desired utility.

\section{Interior point utility maximization for TTL cache hierarchies under random Network Delays}
\label{Sect:Interior point}

In this section, we derive the solution to the utility maximization problem defined in Sect.~\ref{Sect:system model}. 
We first derive a closed form solution for the optimal TTL values for a single cache before solving the utility maximization problem for the cache hierarchy using the interior point method.
\subsection{Straw man: Maximizing the utility of a single cache under exponential distributions}
Next, we derive a closed form solution to the utility maximization of a simple single cache where the mutually independent inter-request times, the TTL and the fetching (network) delays are each independently and identically distributed (iid), specifically, exponentially distributed, with parameters $1/\lambda_i$, $1/\mu_i$ and $1/\mu_F$, respectively. 
Fig.~\ref{Single_cache_MAP_ex1} shows the single cache MAP for an arbitrary object.
The object hit probability and occupancy are obtained from \eqref{eq:hitpr_MAP}-\eqref{eq:Occ_MAP} as
\vspace{-5pt}
\begin{equation}
    P_i=\E[O_i]=\frac{\lambda_i \mu_F}{\mu_i(\mu_F+\lambda_i)+\lambda_i \mu_F }
    \label{eq:MMM single cache performance metrics}
    \vspace{6pt}
\end{equation}
Note that $P_i=\E[O_i]$ is only valid for memoryless inter-request time, TTL and delay distributions.
Now, achieving maximum utility is equivalent to optimizing $P_i$, hence, calculating the corresponding optimal TTL parameter $\mu_i$ in~\eqref{eq:MMM single cache performance metrics}.
%
%
The optimization function \eqref{eq:obj_fn_final}-\eqref{eq:Main_opt_prob_avg_occupancy_constraint} becomes 
\begin{align}
\label{eq: opt_fn MMM single cache}
    \underset{P_i}{\text{maximize}} \  \sum_i^N \lambda_{Ii} \psi(P_{i}) \;, \;
\text{subject to} \sum_i^N P_i=B  \;, 
\end{align}
with $P_i$ as a decision variable for this cache and  $\lambda_{Ii}=\lambda_i$ for the exponentially distributed inter-request times.
The optimization function as expressed in \eqref{eq: opt_fn MMM single cache} coincides with the one proposed in \cite{Deghan_utility} for a single cache with delays. 
Note, however, that a difference exists in the relation between the hit probability and TTL that depends in our single cache model on the delay as expressed within \eqref{eq: opt_fn MMM single cache}.
The following result can also be obtained from~\cite{Deghan_utility}.

The optimal hit probability for the example above is 
$P_i=  \psi'^{-1}(\beta/\lambda_i)$ with 
$\sum_i \psi'^{-1}(\beta/\lambda_i)=B $
where $\psi'^{-1}(\beta/\lambda_i)$ is the inverse function of the derivative of the utility with respect to $P_i$.
Here, $P_i$ is expressed in terms of $\beta$ which is implicitly given and numerically obtained when a given utility function is inserted above. 
For example, in the case of proportional fairness utility ($\alpha=1$) the optimal hit probability is 
$P_i=\frac{\lambda_i}{\sum_i \lambda_i} B$.
Consequently, the optimal TTL for this simple single cache with delay is calculated from \eqref{eq:MMM single cache performance metrics} as
\begin{equation*}
    \hat{\mu_i}=\frac{1}{\mu_F+\lambda_i} \left(\frac{\mu_F \sum_i\lambda_i }{B}-\lambda_i \mu_F\right)    \;. 
\end{equation*}

\subsection{Why maximizing the utility of a cache hierarchy is hard?}


The approach illustrated above of finding the optimal object hit probabilities $P_i$ that maximize the utility and, hence, calculating the corresponding optimal TTL parameters $\mu_i$ is feasible under two conditions: (a) the object hit probability is a bijective function of the TTL parameter (b) the object occupancy function is a composite function $g(P_i(\pmb{\mu}_i))$.

The first condition is only valid for a single cache while the second is only guaranteed in case of a single cache with exponentially distributed iid inter-request times, TTL and delays. 
Extending this simple single cache to PH distributed inter-request times, TTL and delays, condition (b) is not guaranteed to be fulfilled. 
Moreover, for a caching tree model, the hit probability $P_{ij}$ is no longer a bijective function of the TTL parameters as it is controlled by the TTL parameter vector representing multiple caches along a path from a leaf $j$ to the root of the hierarchy.
Trivially, for a leaf cache connected to a parent, storing an object permanently in either cache achieves an object hit probability of one.

Next, we illustrate our approach to the utility maximization problem using the optimal TTLs as decision variables.
Since the problem is a non-linear inequality constrained one, we build on the iterative interior-point approach.

\subsection{Non-linear inequality constrained optimization}
\label{Sect:Interior point_general}
The interior-point optimization method is used to solve non-linear optimization problems of the form of
\begin{align*}
\vspace{-5pt}
\underset{\pmb{x}}{\text{minimize}} \ & \Fx \;,\; \text{subject to} \ \ \Cx=0 \; , \; \pmb{x}>0\;,
\label{eq:general_opt_fn}
\end{align*}
%
where $f: \mathbb{R}^n \rightarrow \mathbb{R}$ is the objective function, and the vector-valued function $c:\mathbb{R}^n \rightarrow \mathbb{R}^m$ contains  $m$ equality constraints.
The interior point approach uses barrier functions to embed the inequality constraints into the objective function~\cite{byrd2000_IP_approach} such that the problem can be rewritten as
\vspace{-5pt}
\begin{align*}
\underset{\pmb{x}}{\text{minimize}} \ & \Fx- \eta \sum_i \ln(x_i)\;, \text{subject to} \ \ \Cx=0 \; ,
\end{align*}
where $x_i$ is the $i$th element of the vector $\pmb{x}$.
The key idea behind the barrier function is to penalize the objective function when approaching the inequality constraint. 
Here, we show the logarithmic barrier function where the penalty of $x_i$ approaching the constraint at $0$ is $-\infty$. 
In addition, the parameter $\eta$ controls the scale of the penalty, where the solution to the problem for a decreasing value of $\eta$ approaches the original problem. 
This requires solving the problem multiple times after reducing the value of $\eta$ gradually. 
At each optimization step (superscript $(i)$), the value of $\eta$ is calculated for the next iteration ensuring a super-linear convergence by~\cite{IPOPT}
\begin{equation*}
\eta^{(i+1)}=\text{max}\left\{\frac{\epsilon_{\text{tol}}}{10}, \text{min}\Big\{\kappa \eta^{(i)}, {\eta^{(i)}}^{\theta} \Big\}\right\}\;,
\vspace{6pt}
\end{equation*}
where $\epsilon_{\text{tol}}$ is the user defined error tolerance, $\kappa \in [0,1]$ and $\theta \in [1,2]$. 
As a result, the value of $\eta$ decreases exponentially as the optimization gets closer to convergence.

For a fixed $\eta$, the optimization problem is solved by applying Newton's method to the Lagrangian
\vspace{-5pt}
\begin{align*}
    \mathcal{L}&=\Fx+\Cx^T\pmb{\nu}-\pmb{x}^T \pmb{z} \;.
    \vspace{8pt}
\end{align*}
Given the Lagrangian multipliers for the equality and inequality constraints, i.e., $\pmb{\nu}$ and $\pmb{z}$, respectively, the method solves the problem iteratively for the steps to the minimum $ \pmb{\delta_x}^{(m)}$, $\pmb{\delta_\nu}^{(m)}$ and $\pmb{\delta_z}^{(m)}$ at iteration $m$ to update the value of the decision variables and the Lagrangian multipliers \cite{IPOPT} according to
\vspace{-3pt}
\begin{align*}
   \pmb{x}^{(m+1)}&=\pmb{x}^{(m)}+\alpha^{(m)} \pmb{\delta_x}^{(m)} \nonumber \\
    \pmb{\nu}^{(m+1)}&=\pmb{\nu}^{(m)}+\alpha^{(m)} \pmb{\delta_\nu}^{(m)} \nonumber \\
    \pmb{z}^{(m+1)}&=\pmb{z}^{(m)}+\alpha_z^{(m)} \pmb{\delta_z}^{(m)} \;.
    \vspace{-3pt}
\end{align*}
where, the step sizes $\alpha$ and $\alpha_z \ \in [0,1]$.
Note that for a more flexible and less restrictive update of the variables, different step sizes are chosen for $\pmb{z}$ than $\pmb{x}$.  
In addition, in order to ensure that the values of $\pmb{x}$ and $\pmb{z}$ are positive at each iteration~\cite{IPOPT}, $\alpha^{(m)}$ and $\alpha_z^{(m)}$ are calculated with respect to a fraction to the boundary value $\varepsilon^{(i)}:=\text{max}\{\varepsilon_{\text{min}},1-\eta^{(i)}\}$ as
\begin{align}
\label{eq:stepsizes_bound}
    \alpha^{(m)}&\leq\text{max} \{\alpha \in [0,1]: \pmb{x}^{(m)}+\alpha \pmb{\delta_x}^{(m)} \geq (1-\varepsilon)\pmb{x}^{(m)} \} \nonumber
    \\
    \alpha_z^{(m)}&=\text{max} \{\alpha \in [0,1]: \pmb{z}^{(m)}+\alpha \pmb{\delta_z}^{(m)} \geq (1-\varepsilon)\pmb{z}^{(m)} \} \;,
\end{align}
where $\varepsilon_{\text{min}} \in [0,1]$.
Furthermore, choosing a specific value of $\alpha^{(m)} \in [0, \alpha^{(m)}_\text{max}]$ from \eqref{eq:stepsizes_bound} by applying a backtracking line search method ensures sufficient progress towards the optimal solution.
Using a variant of Fletcher and Leyffer’s filter method~\cite{Line_search_SQP}, the authors of \cite{Line_search_IPOPT} proved under mild conditions that the interior point achieves global convergence. 

The optimal steps $ \pmb{\delta_x}^{(m)}$, $\pmb{\delta_\nu}^{(m)}$ and $\pmb{\delta_z}^{(m)}$ at iteration $m$ are calculated by solving 
{\small
\begin{equation}
\setlength\arraycolsep{2pt}
    \begin{bmatrix}
\mathbf{H}^{(m)} & \mathbf{J}^{(m)} & -\mathbf{I} \\
{\mathbf{J}^{(m)}}^T & \pmb{0} &         \pmb{0}       \\
\mathbf{Z}^{(m)} & \pmb{0} &         \mathbf{X}^{(m)}
\end{bmatrix}
\begin{bmatrix}
\pmb{\delta_x}^{(m)}  \\
\pmb{\delta_\nu}^{(m)}  \\
\pmb{\delta_z}^{(m)}
\end{bmatrix}
= -
\begin{bmatrix}
 \mathbf{\Gamma} \\
{c}(\pmb{x}^{(m)})      \\
\mathbf{X}^{(m)}\mathbf{Z}^{(m)}\pmb{1}-\eta\pmb{1} 
\end{bmatrix}
\;,
\label{eq:Newtons_method}
\vspace{3pt}
\end{equation}
}
\noindent
where $\mathbf{H}$ is the Hessian of the Lagrangian $\mathcal{L}$ defined as
\begin{equation*}
  \mathbf{H}:=\nabla^2_{xx}\mathcal{L}=\nabla_{xx} \Fx+  \sum_k \nu_k \nabla_{xx} c_k(\pmb{x})\;
\end{equation*}
and using the shorthand matrix $\mathbf{\Gamma}:=\nabla f(\pmb{x}^{(m)})+\mathbf{J}^{(m)} {\pmb{\nu}^{(m)}}-\pmb{z}^{(m)}$, 
where $\mathbf{J}:=~\nabla \Cx$ is the Jacobian matrix of the equality constraints. Here, 
$\mathbf{X}^{(m)}$ and $\mathbf{Z}^{(m)}$ are diagonal matrices of $\pmb{x}^{(m)}$ and $\pmb{z}^{(m)}$, respectively.
\vspace{-5pt}
\subsection{Utility maximization for TTL cache hierarchies}
\label{Sect:Interior point_derivation}
Next, we first apply the interior point optimization to solve the TTL cache utility maximization~\eqref{eq:obj_fn_final}-\eqref{eq:Main_opt_prob_avg_occupancy_constraint}.
We express the objective function $\Fx$ and the constraints $c_k(\pmb{x})$ as
\begin{align}
    \Fx&=- \sum_i \sum_j \lambda_{Iij} \psi(P_{ij}(\pmb{\mu}_i)) \nonumber \\
    \vspace{-5pt}
    c_k(\pmb{x})&= \sum_i \E[O_i^k] -B_k, \ \forall k \in \{0,..,n_c\}
    \vspace{-5pt}
    \label{eq:utility_max_interior_point}
\end{align}
The decision column vector $\pmb{x}$ contains \textit{the TTL parameters of each object at each cache} and controls the aggregate utility of the hierarchy.
We express $\pmb{x}$ in terms of the TTL parameters of each object as 
\begin{equation*}
    \pmb{x}=[\pmb{\mu}_1^T,\ \pmb{\mu}_2^T, \dots, \pmb{\mu}_N^T]^T
    \vspace{6pt}
\end{equation*}

Next, we derive the gradients $\nabla_{xx} \Fx$ , $\nabla \Fx$,  $\nabla_{xx} \Cx$ and $\nabla \Cx$ for the cache utility maximization problem to calculate $\H$, $\J$ and $\bGamma$ from \eqref{eq:Newtons_method}, to obtain the optimal search directions.
We formulate the gradient and Hessian of $\Fx$ with respect to the TTL parameters of each object $\pmb{\mu}_i$. 
The key to using interior point optimization is that TTL caching decouples the objects in the cache, i.e, the hit probability of an object $i$ only depends on its TTL parameters $\pmb{\mu}_i$, and \textit{thus we can express $\nabla \Fx$ and $\nabla_{xx} \Fx$ as}
\begin{equation}
\nabla \Fx=
    \begin{bmatrix}
        \nabla^T_{\pmb{\mu}_1} \Fx & \nabla^T_{\pmb{\mu}_2} \Fx   &
        \dots &   \nabla^T_{\pmb{\mu}_N} \Fx 
    \end{bmatrix}^T
    \label{eq:grad_f}
\end{equation}

\begin{equation}
\nabla_{xx} \Fx=
    \begin{bmatrix}
       \nabla_{\pmb{\mu}_1\pmb{\mu}_1} \Fx & \pmb{0} & \pmb{0} \\
         \pmb{0} &\ddots& \pmb{0}\\
       \pmb{0} &  \pmb{0}&\nabla_{\pmb{\mu}_N\pmb{\mu}_N}\Fx  
    \end{bmatrix}
    \label{eq:hess_f}
    \vspace{6pt}
\end{equation}
Similarly, $\nabla \Cx$ and $\nabla_{xx} \Cx$ are represented as in \eqref{eq:grad_f} and \eqref{eq:hess_f}, which we do not show here for space reasons. 

Recall that the object hit probability \eqref{eq:hitpr_MAP_perleaf} and the expected occupancy \eqref{eq:Occ_MAP} are calculated using the steady state probabilities $\pmb{\pi}_i$, which in turn depend on the transition matrices of the hierarchy MAP.  
Therefore, we next derive $\nabla \Fx$, $\nabla_{xx} \Fx$, $\nabla \Cx$ and $\nabla_{xx} \Cx$ in terms of the derivatives of the steady state vector of each object $i$ with respect to its TTL parameters. 

Given the TTL parameters $a,b$ each representing any of the TTL parameters in $\pmb{\mu}_i$ of an object $i$. From \eqref{eq:steady_state_calc} the derivatives $\pmb{\pi}^{a'}_{i}:=\frac{\partial  \pmb{\pi}_{i}}{\partial a}$ and  $\pmb{\pi}^{a' b'}_{i}:=\frac{\partial^2 \pmb{\pi}_{i}}{\partial a \partial b}$ are given as
\begin{align}
\vspace{-10pt}
    \pmb{\pi}^{a'}_{i} &=-\pmb{b} \mathbf{A}^{-1}_{i} \mathbf{A}^{a'}_{i} \mathbf{A}^{-1}_{i} 
    \nonumber
    \\
    \pmb{\pi}^{a' b'}_{i}
       &=\pmb{\pi}_i[ \mathbf{A}^{a'}_{i} \mathbf{A}^{-1}_{i} \mathbf{A}^{b'}_{i}  + \mathbf{A}^{b'}_{i}  \mathbf{A}^{-1}_{i}   \mathbf{A}^{a'}_{i}  ]\mathbf{A}^{-1}_{i}
       \label{eq:diff std state}
       \vspace{-5pt}
\end{align}
\begin{theorem}
\label{theo:Diff_Fx}
The derivatives $f^{a'}:=\frac{\partial \Fx}{\partial a}$ and  $f^{a'b'}:=\frac{\partial^2\Fx}{\partial a \partial b}$ associated with the TTL cache objective function \eqref{eq:obj_fn_final} are given in terms of the derivatives of the steady state vector $\pmb{\pi}^{a'}_{i}$ and $\pmb{\pi}^{a'b' }_{i}$ from~\eqref{eq:diff std state} as
\begin{align} 
 f^{a'}&=\sum_j\ \psi' \pmb{\pi}^{a'}_{i} \mathbf{D}_{{ij,1}} \pmb{1} \;, 
    \label{eq:obj_diff} \\
    \vspace{-10pt}
    f^{a' b'} &=
    \sum_j 
    \left[ \lambda_{Iij} \psi' \pmb{\pi}^{a'b' }_{i}-\psi'' \pmb{\pi}^{b'}_{i}
    \mathbf{D}_{{ij,1}} \pmb{1} \pmb{\pi}^{a'}_{i}   
    \right]
    \frac{\mathbf{D}_{{ij,1}} \pmb{1}}{\lambda_{Iij}} 
    \label{eq:obj_double_diff}
    \vspace{-5pt}
\end{align}
where $\psi'$ is the derivative of $\psi$ with respect to $P_{ij}$.
\end{theorem}
\noindent The proof of Thm.~\ref{theo:Diff_Fx} is in the appendix.

Now, the equality constraint is expressed in terms of the steady state vector according to \eqref{eq:Occ_MAP}.
Therefore, we calculate its first and double differentiation as
\begin{align}
\vspace{-5pt}
c^{a'}_k(\pmb{x})&= \frac{\partial \E[O_i^k]}{\partial a}=  \sum_{l \in \chi_i^k} {\pi}^{a' }_{{i},l}\;, \nonumber \\[-0.5em]
c^{a'b'}_k(\pmb{x})&= \frac{\partial^2 \E[O_i^k]}{\partial a \partial b}= \sum_{l \in \chi_i^k} {\pi}^{a' b' }_{{i},l} \;,
\label{eq:constraint_diff}
\vspace{-5pt}
\end{align}
where ${\pi}^{a' }_{{i},l}$ and  ${\pi}^{a' b'}_{{i},l}$ are the $l$-th elements of  $\pmb{\pi}^{a' }_{i}$ and $\pmb{\pi}^{a' b'}_{i}$, respectively.
We can now use~\eqref{eq:diff std state} and Thm.~\ref{theo:Diff_Fx} to calculate $\nabla_{x} \Fx$ and $\nabla_{xx} \Fx$.
Similarly, we use~\eqref{eq:diff std state},~\eqref{eq:constraint_diff} to calculate $\nabla_{xx} \Cx$ and $\nabla_x \Cx$.
As a result, we obtain $\H$, $\J$ and $\bGamma$.  

\section{Storage considerations for TTL optimization}
\label{Sect:meTTL}
Next, we discuss the practical limitations of using the expected occupancy in the constraint~\eqref{eq:Main_opt_prob_avg_occupancy_constraint} and propose an empirical extension to circumvent these limitations.

\vspace{-5pt}
\subsection{Cache storage size violation}
Given abundant cache storage, TTL assignment and TTL-based eviction treat objects independently.
This allows the TTL policy to imitate different caching policies by tuning the TTLs to generate the same \textit{expected} object occupancy and average performance as under other policies such as LRU~\cite{Che:LRUtoTTL, Fofack:TTL}.

The constraint~\eqref{eq:Main_opt_prob_avg_occupancy_constraint} links the average occupancy of the objects to the cache size, which guarantees the full utilization of the cache storage on average as the average number of cached objects equals the (hypothetical) cache size $B_k$.
The drawback of this consideration is that the number of cached objects is randomly distributed while centered around $B_k$~\cite{Fricker:2012:versatile}.
Any caching system has a finite storage capacity, hence, the cache $k$ will use a hard eviction policy to evict an object when $B_k$ objects are present upon object insertion. 
Hence, we require a decision rule in addition to the optimal TTLs obtained from the closed-form result from Sect.~\ref{Sect:Interior point} to evict an object when the cache is full and an object has to be admitted.
To this end, we propose that when an object miss occurs at a cache $k$ which already includes $B_k$ objects, we use the minimum TTL policy (denoted \textbf{TTL$_{\min}$}) to evict the object which has the minimum remaining lifetime (TTL).
Consequently, the object will not be admitted if its fresh TTL is smaller than the remaining lifetime of all the objects in the cache.
As a result of the eviction of objects prior to their optimal TTL values, this extension achieves a lower utility than the one obtained through analytical optimization.


For a single cache,~\cite{Deghan_utility} shows that the storage size violation probability approaches zero in a scaling limit.
For a finite size system, the achieved utility is, however, not optimal. 


\subsection{Heuristic: Minimum extended Time-to-Live (TTL$_{\min,\text{extnd}}$) }
In order to counteract the decrease in utility due to finite storage size we additionally propose a heuristic TTL-based policy, which we denote as minimum extended Time-to-Live (TTL$_{\min,\text{extnd}}$).
TTL$_{\min,\text{extnd}}$ leaves the object in the cache despite TTL expiry when the cache is not full which reduces the utility loss from underutilizing the cache storage.

We calculate the TTL-based timestamp of eviction of an object $i$ in the cache at time t as
\vspace{-5pt}
\begin{equation}
    \tau_i(t):=\hat{\tau_i}-t+t_i', \ \ \tau_i,t, t_i' \in \mathbb{R}, \ t\geq t_i' \;,
    \label{eq:tag_ttl_min_extnd}
    \vspace{4pt}
\end{equation}
where $t_i'$ is the latest TTL resetting time of object $i$ either from a hit or a new admission to the cache. We denote $\hat{\tau_i}$ as the analytically optimized TTL value, thus initially $\tau_i(t_i')=\hat{\tau_i}$. 
A negative value of $\tau_i(t)$ represents the extended time to live of a delayed eviction.

In steady state, using TTL$_{\min,\text{extnd}}$ we evict an object only upon admission of another object to a full cache.
In addition, this heuristic maintains the desired utility by evicting the objects based on the optimally calculated TTL values. 
Upon a request to a full cache at time $t$, TTL$_{\min,\text{extnd}}$ evicts the object with minimum TTL $ \tau_i(t)$ from \eqref{eq:tag_ttl_min_extnd} expressed as $\hat{i}=\underset{i  \in  \mathbb{O}(t)}{\text{argmin}} \ \tau_i(t)$,
where $\hat{i}$ is the tag of the object with minimum TTL and $\mathbb{O}(t)$ is the set containing all the objects in the cache at $t$ including the object to be admitted. 
Note that the rule above will let the fresh object only be admitted if its TTL is at larger than that of an object in the cache.
This rule implies that, on the one hand, the early eviction of an object with $\tau_i(t)>0$, i.e., before the expiry of the optimized TTL, is possible.
On the other hand, the rule implies that a delayed eviction of an object with negative $\tau_i$, i.e. an object with the earliest expired TTL \textit{preserves the ordering specified by the optimal TTL values.}
We show by simulations in Sect.~\ref{Sect:evaluations} that the utility of TTL$_{\min,\text{extnd}}$ approaches the analytical one.

\section{Learning on Graph transformations of  TTL Cache hierarchies}
\label{Sect:GNN}
In contrast to the closed-form solution presented in Sect.~\ref{Sect:Interior point}, we propose next a learning-based approach to predict the optimal TTLs.
Our rationale for using \acp{GNN} is that these can process graphs of large sizes, circumventing the scaling limitation of the closed-form optimization.
We evaluate the execution time of both methods in Sect.~\ref{Sect:evaluations}.

Next, we present our neural network architecture and data transformation used for optimizing the TTL configuration of the caching hierarchy.
The goal is to enable an agent to process the caching hierarchy and its properties (e.g. structure, cache capacities, request arrival rates) and predict the optimal TTL to use for each object at each cache.
Our approach is based on a transformation of the caching hierarchy into a graph data structure, which is then processed using a \ac{GNN}.

\subsection{A GNN approach}
We use the framework of \acp{GNN} introduced in~\cite{Gori2005,Scarselli2009}.
These are a special class of \acp{NN} for processing graphs and predicting values for nodes or edges depending on the connections between nodes and their properties.
Fundamentally, \acp{GNN} utilize \emph{message passing} where messages are updated and passed between neighboring nodes. Essentially, these messages are vectors  $\B{h}_v \in \mathbb{R}^k$ (here $k=2^7$) that are propagated throughout the graph over multiple iterations.
We refer to \cite{Bronstein2021} for a formalization of \ac{GNN} concepts.

We select \ac{GGNN}~\cite{Li2016a} as \ac{GNN} model, with the addition of edge attention.
For each node $v$ in the graph, its message $\B{h}_v^{(t)}$ is updated at iteration $t$ as
\begin{align}
	\B{h}_v^{(t=0)} & = \mathit{FFNN}_{\mathit{init}}\left(\iota_v \right)  \label{eq:gnn_init} \\
	\B{h}_v^{(t)} &  = \mathit{GRU}\bigg( \B{h}_{v}^{(t-1)}, \sum_{u \in q(v)} \lambda_{(u, v)}^{(t-1)} \B{h}_{u}^{(t-1)} \bigg) \label{eq:gnn_f_sum} \\
	\lambda_{(u, v)}^{(t)} & =  \sigma \left( \mathit{FFNN}_{\mathit{edge}}\left( \left\{ \B{h}_{u}^{(t)}, \B{h}_{v}^{(t)} \right\} \right) \right) \label{eq:edge_attention} \\
	\text{o} & = \mathit{FFNN}_{\mathit{out}}(\B{h}_v^{(d)}) \label{eq:gnn_output}
\vspace{4pt}
\end{align}
with $\iota_v$ and $\text{o}$ being the input at node $v$ and the output, respectively. Also $\sigma(x) = 1 / (1 + e^{-x})$ denotes the sigmoid function,
$q(v)$ the set of neighbors of node $v$,
$\{a, b\}$ the concatenation operator of vectors $a$ and $b$, 
$\mathit{GRU}$ a \ac{GRU} cell, 
$\mathit{FFNN}_{\mathit{(\cdot)}}$ are \acp{FFNN},
and $\lambda_{(u, v)} \in (0,1)$ being the weight for the edge $(u, v)$.
We obtain the final prediction for each node by a \ac{FFNN} \eqref{eq:gnn_output} after applying \eqref{eq:gnn_f_sum} for $d$ iterations, with $d$ corresponding to the diameter of the analyzed graph.

\subsection{Graph transformation}
Next, we use the graph induced through the hierarchy, where nodes are caches and edges are links, as an undirected graph data-structure that is processed using the \ac{GNN}.
%
Each node has an input vector of fixed size describing its
features.
Specifically, we use the (i) cache type, i.e. leaf or non-leaf node, (ii) average network delay to the parent cache, and (iii) cache capacity.
Additionally, the following features are used for the leaves: (iv) sum of expected inter-arrival time for all the objects, (v) expected inter-arrival time of requests for the top-$M$ objects, and (vi) expected inter-arrival time for the remaining $M+i$ objects.
For each node, an output vector of fixed size is predicted, with the following features: (a) $M$ TTL values to use for the top-$M$ objects, (b) A default TTL value for the remaining $M+i$ objects, (c) the index of the node where each object should be cached in the hierarchy.
The last restriction stems from the efficiency goal to cache an object only at a single cache along each leaf-root path. 





\subsection{Training approach}
In order to train the \ac{GNN}, we use \ac{RL} using the \texttt{REINFORCE} policy gradient algorithm~\cite{williams1992_RL} where we use a basic training loop, where the TTL configuration and the object locations predicted by the \ac{GNN} are used as input configuration for a cache simulator.
At the end of a simulation run, the empirical object hit probabilities are computed and fed back to the training loop as reward.

The reward function depends on the chosen fairness utility function from Sect.~\ref{sec:fairness}.
Note that we enrich the reward vector with the utility for each object in the cache and in the case of the log utility function in~\eqref{eq:alphafair_utility} we obtain its value close to zero using its Taylor approximation of order 30. 




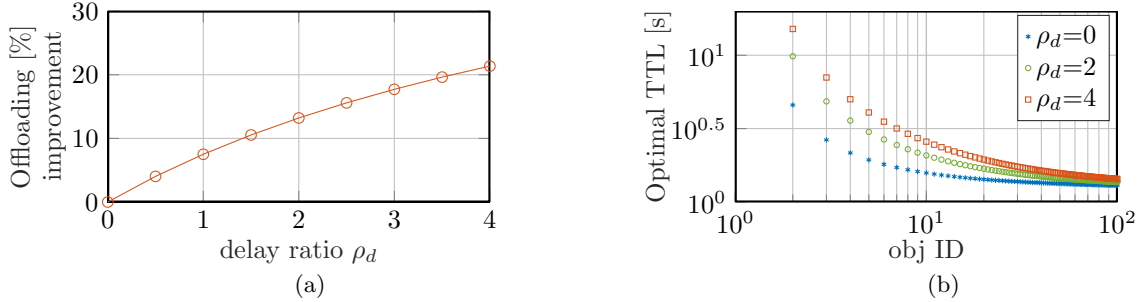
\begin{figure}[t]
  \captionsetup[subfigure]{margin={1.4cm,0cm}}

     \centering
     \begin{subfigure}[b]{0.49\textwidth}
         \centering
	{\input{Evaluations/Delay_Impact_single_cache/Throughput_Occ_errors_arxiv} }
     \captionsetup{skip=0pt} 
	\caption{}
	\label{fig:delay_imp_single_D}
     \end{subfigure}
     \hfill
          \begin{subfigure}[b]{0.49 \textwidth}
         \centering
	{\input{Evaluations/Delay_Impact_single_cache/TTL_values_arxiv.tex}}
     \captionsetup{skip=0pt} 
	\caption{}
	\label{fig:delay_imp_single_C}
     \end{subfigure}
     \caption{\textbf{Single cache results:} TTL optimization under random network delay vs. the idealized zero-delay assumption where $\rho_d$ denotes the ratio of the expected network delay to the expected inter-request time of the hottest object.
    (a) Significant Improvement of origin offloading, i.e. traffic served by the cache, due to optimizing under delays (OPT$|_{\text{delay}}$) vs. idealization (OPT$|_{\text{ideal}}$) increases with the delay.
    (b)~The optimal TTL values are larger for OPT$|_{\text{delay}}$ especially for hottest objects. Object IDs sorted in descending arrival rate.
        }
     \label{fig: delay_imp_single}
     \vspace{-15pt}
\end{figure}

\section{Evaluations and Lessons learned}
\label{Sect:evaluations}
Next, we show numerical performance evaluations of the proposed cache utility maximization approaches. 
If not stated otherwise we consider the two-level caching tree in Fig.~\ref{fig:caching_tree} and use the proportional fairness parametrization ($\alpha=1$) of the utility function~\eqref{eq:alphafair_utility}.
The request rates at the leaf caches follow a Zipf distribution, i.e., the rate of index $j$ is $\lambda_j=\frac{1}{j^s}$ and $s=0.8$ as observed in~\cite{Fricker:2012:versatile}. 
Note that for a given object the request rates are inhomogeneous across the leaves, i.e., an object $i$ is assigned an index $j$ at each leaf uniformly at random.
Homogeneity follows accordingly.
Empirical values stem from long enough emulations of at least $5 \times 10^5$ requests to generate enough events for cold objects.
For illustration, we let the inter-request times, network delays and TTLs to be exponentially distributed and illustrate the impact of the network delays on the optimal utility. 
As noted in Sect.~\ref{sec:primer_TTL_exact} it is simple to include these three model components as MAPs into Thm.~\ref{theo:Diff_Fx}.
We use Pytorch~\cite{Pytorch} to train the GNN and the evaluations run on a 24-core machine with $ \qty{64}{\giga\byte}$ RAM. 
\subsection{Single cache optimization under random network delays}

First, we consider a single cache under random network delays to the origin server. 
We use $N=100$ objects under Zipf-popularity and capacity $B=10$.
Fig.~\ref{fig:delay_imp_single_D} shows a strong improvement in a classical cache metric, i.e., the origin offloading, when optimizing the object TTLs considering the random network delay compared to ignoring it.
The origin offloading is the fraction of request traffic served by the cache and it is obtained from the utility~\eqref{eq:alphafair_utility} by $\alpha=0$.
Here, we vary the ratio $\rho_d$ of the expected network delay to the expected inter-request time of the hottest object based on the arguments in~\cite{AtreSWB20,Elsayed_cache_hierarchy_MAP}. 
One manifestation of the observed improvement are the \textit{per-object optimal TTLs} that are higher when accounting for the random delays as shown in Fig.~\ref{fig:delay_imp_single_C}.
Observe that the possible loss in hit probability, consequently utility, due to misses during the object fetch delay is compensated by the longer TTLs.

\subsection{Hierarchy optimization under random network delays}
\begin{figure}[t]
  \captionsetup[subfigure]{margin={1.2cm,0cm}}
     \centering
     \begin{subfigure}[b]{0.49\textwidth}
         \centering
	{\input{Evaluations/Delay_Impact_on_three_caches/Hierarchy_vs_individual_arxiv}}
     \captionsetup{skip=0pt} 
	\caption{}	\label{fig:Hier_individ_utility}
     \end{subfigure}
           \hfill
     \begin{subfigure}[b]{0.49\textwidth}
         \centering
	{\input{Evaluations/Delay_Impact_on_three_caches/Error_EKAD_arxiv}} 
     \captionsetup{skip=0pt} 
	\caption{}
	\label{fig: pract_policies}
     \end{subfigure}
     \caption{\textbf{Two-level binary tree:} (a)~Aggregate utility of OPT$|_{\text{delay}}$ vs.OPT$|_{\text{ideal}}$ vs. LRU hierarchy. 
     (b)~Utility loss under strict cache capacity constraint with respect to the optimal utility without strict capacity restriction \eqref{eq:Main_opt_prob_avg_occupancy_constraint}. TTL$_{\min}$ has a $10\%$ deviation due to the cache storage violation while  the heuristic TTL$_{\min,\text{extnd}}$ approaches the optimal utility.}
     \label{fig: delay_imp_single}
\end{figure}
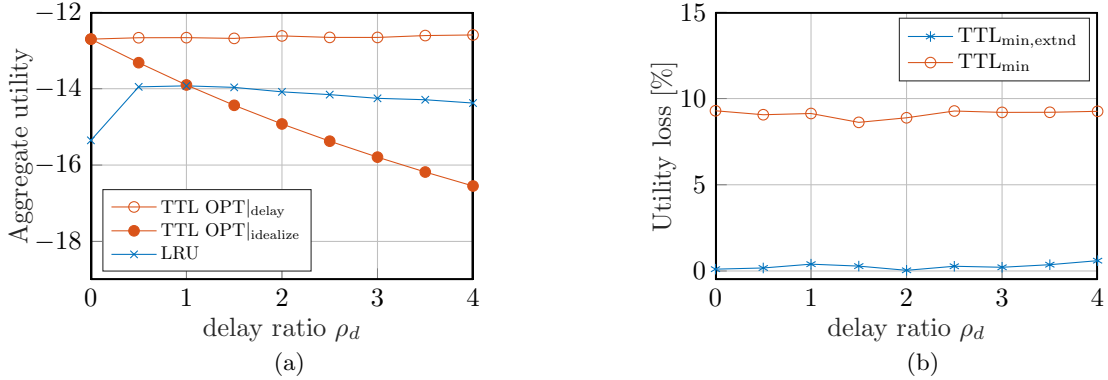

Next, we return to the example in Fig.~\ref{fig:Obj_hitpr_delrat4_intro}  and show the influence of optimizing the hierarchy caches together compared to using traditional algorithms such as LRU throughout the hierarchy. We parameterized the hierarchy with $N=100$ objects and a storage of $B=5$ per cache. 
Fig.~\ref{fig:Hier_individ_utility} shows 
that in contrast to the detrimental impact of the increasing delay on the utility of OPT$|_{\text{ideal}}$ 
we can maintain the optimal utility for varying $\rho_d$ using the optimization from Sec.~\ref{Sect:Interior point_derivation} (denoted OPT$|_{\text{delay}}$).
Observe that the aggregate utility remains higher than that of LRU which is utility agnostic.

\subsection{Cache size restrictions and empirical  TTL policies}
Next, we consider \textit{emulations} of the cache hierarchy under the same parameterization as above where we use the closed-form optimized TTLs obtained under the average occupancy constraint~\eqref{eq:Main_opt_prob_avg_occupancy_constraint} as in Sect.~~\ref{Sect:Interior point}. 
In Fig.~\ref{fig: pract_policies}, we compare the relative utility loss of using these optimized TTLs derived using~\eqref{eq:Main_opt_prob_avg_occupancy_constraint}, when used together with the policies TTL$_{\min}$, TTL$_{\min,\text{extnd}}$ from Sect.~\ref{Sect:meTTL} under a strict capacity constraint for an increasing delay ratio~$\rho_d$.
This utility loss is with respect to the analytical optimal aggregate utility, i.e. OPT$|_{\text{delay}}$, that uses~\eqref{eq:Main_opt_prob_avg_occupancy_constraint}.
The figure shows an almost $10\%$ deviation from the analytically obtained optimal utility when applying the \textbf{TTL$_{\min}$} policy (i.e. evicting the object with the minimum TTL upon exhausting the cache capacity) highlighting the effect of the cache storage violation.
Further, by allowing the TTL extension when the cache is not full, \textbf{TTL}$_{\min,\text{extnd}}$ remarkably shows less than $1\%$ deviation from the optimal utility. We note, however, that this policy is hard to evaluate analytically.

\subsection{Scaling with the hierarchy size: A GNN to the rescue}
\begin{figure}[t]
\centering
      \begin{subfigure}[b]{1\linewidth}
            \centering
     \resizebox{0.5\linewidth}{!}{\input{Evaluations/Run_time.tex}}
        \label{}
     \end{subfigure}
         \\
     \vspace{2pt}
    \begin{subfigure}[b]{1\linewidth}
            \centering
                 \hspace{-0.04cm}
         \resizebox{0.5\linewidth}{!}{\input{Evaluations/Acc.tex}}
        \label{}
     \end{subfigure} 
          \captionsetup{skip=0pt} 
\caption{\textbf{Scaling the number of caches:} Comparison of the optimal aggregate utility and the execution time of the analytical and the GNN-based TTL optimization. OPT$|_{\text{delay}}$ becomes intractable while the GNN provides good results at a small execution time.}
    \label{fig:GNN_result_hierarchy}
\end{figure}
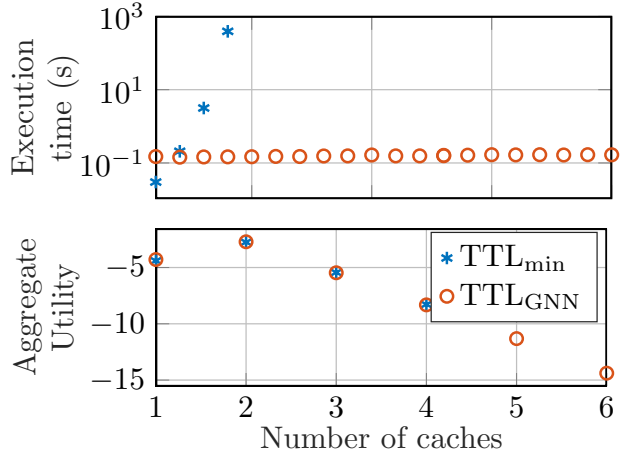

As we increase the number of leaf caches in the two level caching tree (cf. Fig.~\ref{fig:caching_tree}) we observe in Fig.~\ref{fig:GNN_result_hierarchy} that the computation time of the optimal TTLs increases exponentially in the number of caches (cf. the discussion in Sect.~\ref{sec:primer_TTL_exact}).
We, hence, show that utilizing the GNN-based approach from Sect.~\ref{Sect:GNN} we obtain very good results while maintaining a constantly minuscule execution time.

Fig.~\ref{fig:GNN_result_hierarchy} compares the performance of cache hierarchy emulations using the GNN approach and using the analytical TTL optimization, specifically  $\textbf{TTL}_{\min}$ for $N=30$ number of objects and a capacity of $B=4$ for each cache.
Note that this comparison is consistent as the GNN is trained on the same cache hierarchy.
We increase the total number of caches $n_c$, where $n_c=1$ represents a single cache and $n_c=2$ denotes one parent and one child cache.
This explains the non-monotonic behavior of the aggregate utility with an increasing number of caches.
The figure shows that the GNN is able to deliver good performance (albeit not analytically proven) at a constantly small execution time.

\subsection{Trace-based evaluation}
\begin{figure}[t]
\centering
      \begin{subfigure}[b]{1\linewidth}
            \centering
     \resizebox{0.6\linewidth}{!}{{\input{Evaluations/Trace/Trace_offloading}}}
        \label{}
     \end{subfigure} 
     \captionsetup{skip=0pt} 
\caption{\textbf{Trace-based evaluation}: Origin offloading of the optimal TTL under strict capacity constraint (TTL$|_\text{{min}}$) relative to LRU, FIFO and Random caching for increasing delay.}
    \label{fig:trace}
\end{figure}
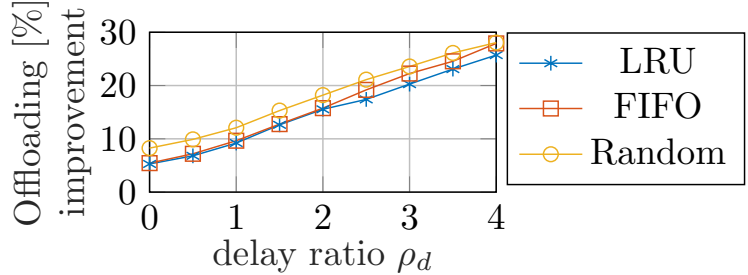

Next, we describe a trace-based evaluation of the presented TTL optimization procedure. 
We use a data center trace~\cite{SNIA_Trace_IBM} having $10^6$ requests to more than $8\times 10^4$ objects as input to a single cache. 
We optimize the TTL cache proportional fairness utility ($\alpha=1)$. 
As the trace contains cold transient objects, 
we only consider for the TTL optimization the most frequent objects receiving each at least 15 requests (overall 580 objects) using the analytical method in Sect.~\ref{Sect:Interior point} and do not cache the colder objects (set their TTLs to zero).
We estimate and verify the mean request rates of each object assuming exponentially distributed inter-request times. 
As the objects may, however, receive requests only for a certain duration within the trace, we use 
the average number of requests per object $\omega_i$ to gauge~\eqref{eq: opt_fn MMM single cache} as $\sum_i^N \sum_j^{n_l} \omega_{i} \psi(P_{i}(\pmb{\mu}_i))$.  
We run simulations using the optimal TTLs vs. vanilla LRU, FIFO and Random caching for a fixed cache size of 50 objects and vary the expected network delay relative to the mean inter-request time.
Fig.~\ref{fig:trace} shows the strong improvement in the amount of traffic served by the cache (origin offloading) due to optimizing the object TTLs under random network delays compared to vanilla LRU, FIFO and Random caching.




\section{Conclusion}
\label{Sect:concl}
We study the optimization of TTL cache hierarchies under random network delays. By leveraging object decoupling in the exact TTL cache model we analytically solve the non-linear utility maximization problem for cache hierarchies and find the optimal per-object TTL values.
As the optimal solution inherits tractability issues from the exact model, we propose a GNN-based approach that scales with the hierarchy size. Numerical and trace-based evaluations of both methods show strong caching performance improvements when incorporating the network delay into the cache hierarchy optimization. 


\section*{Appendix}
\label{Sect:appendix}
\vspace{-5pt}
\subsection{Proof of Theorem~\ref{theo:Diff_Fx}}
\label{app:diff_proof}
First, we prove \eqref{eq:obj_diff}.
Let $a$ represent any of the TTL parameters in $\pmb{\mu}_k$ of object $k$.
The partial derivative of $f$ with respect to $a$ is represented as
\vspace{-5pt}
\begin{equation}
        f^{a'}= -\frac{\partial}{\partial a} \bigg(\sum_i^N \sum_j^{n_l} \lambda_{Iij} \psi(P_{ij}) \bigg) \ .
\end{equation}  
All the terms of the first summation do not depend on $\mu_k$ except for the term at $i=k$. Therefore,
\begin{equation}
       f^{a'}=-\sum_j^{n_l} \lambda_{Ikj} \frac{\partial \psi(P_{kj})}{\partial a}=-\sum_j^{n_l} \lambda_{Ikj} \frac{\partial \psi}{\partial P_{kj}} \frac{\partial P_{kj}}{\partial a} \ .
    \label{eq:obj_fn_diff}
\end{equation}  
$\psi'\coloneqq \frac{\partial \psi}{\partial P_{kj}}$ is calculated depending on the chosen utility notion, e.g., for proportional fairness utility, $\psi'~=~[\ln(10) \log(P_{kj})]^{-1}$. 
Using \eqref{eq:hitpr_MAP_perleaf}, we obtain $\frac{\partial P_{kj}}{\partial a}$ in terms of $\pmb{\pi}^{a'}_{{k}}=\frac{\partial \pmb{\pi}_{{k}}}{\partial a}$ as 
$    \frac{\partial P_{kj}}{\partial a}=- \frac{1}{\lambda_{Ikj}} \pmb{\pi}^{a'}_{{k}}  \mathbf{D}_{{kj,1}} \pmb{1} 
    \label{eq:hit_pr_diff}
$.
Note that $\mathbf{D}_{{kj,1}}$ only depends on the request process. 
Using $\frac{\partial P_{kj}}{\partial a}$ in \eqref{eq:obj_fn_diff}, we obtain \eqref{eq:obj_diff}.
Next, we prove \eqref{eq:obj_double_diff}. We derive
\begin{align*}
     f^{a' b'}&=\frac{\partial f^{a'}}{\partial b}  =\sum_j\ \frac{\partial (\psi' \pmb{\pi}^{a'}_{{k}} )}{\partial b} \mathbf{D}_{{kj,1}} \pmb{1} 
     \nonumber \\
     &=\sum_j \big[\psi' \pmb{\pi}^{a'b'}_{{k}}+ \psi''  \frac{\partial P_{kj}}{\partial b}  \pmb{\pi}^{a'}_{{k}}\big] \mathbf{D}_{{kj,1}} \pmb{1} \ .
\end{align*}
Inserting the expression of $\frac{\partial P_{kj}}{\partial b}$ from above we obtain 
\begin{equation*}
     f^{a' b'} =\sum_j \big[\lambda_{Ikj} \psi' \pmb{\pi}^{a'b'}_{{k}}-  \psi''   \pmb{\pi}^{b'}_{{k}}  \mathbf{D}_{{kj,1}} \pmb{1}   \pmb{\pi}^{a'}_{{k}} \big] \frac{\mathbf{D}_{{kj,1}} \pmb{1} }{\lambda_{Ikj}} \;.
\end{equation*}


\balance
\bibliographystyle{IEEEtran.bst}
\bibliography{IEEEabrv,main.bib}

\end{document}

%% file: Evaluations/Delay_Impact_on_three_caches/Motivation.tex
%
%
\definecolor{mycolor1}{rgb}{0.00000,0.44700,0.74100}%
\definecolor{mycolor2}{rgb}{0.85000,0.32500,0.09800}%
\definecolor{mycolor3}{rgb}{0.2,0.6,0.4}%

\begin{tikzpicture}

\begin{axis}[%
width=1.7in,
height=0.8in,
scale only axis,
xmin=0,
xmax=3,
xlabel style={font=\color{white!15!black},font=\Large},
xlabel={$\ \frac{\mathsf{E[}\text{network delay}]}{\mathsf{E[}\text{inter-request time}]}$},
ymin=-16,
ymax=-12,
ylabel style={font=\color{white!15!black}},
ylabel={Aggregate utility},
axis background/.style={fill=white},
ylabel style={yshift=-0.3cm},
xlabel style={yshift=0.1cm},
xlabel style={yshift=0.15cm},
axis x line*=bottom,
axis y line*=left,
xmajorgrids,
ymajorgrids,
legend style={legend cell align=left, align=left, draw=white!15!black},
legend style={nodes={scale=0.8}},
legend pos=outer north east
]
\draw[line width=0.002 \linewidth](current axis.south west)rectangle(current axis.north east);


\addplot [color=mycolor2,mark=o]
  table[row sep=crcr]{%
1e-05	-12.7017615574456\\
0.5	-13.3187957438792\\
1	-13.9019533961636\\
1.5	-14.4350490866435\\
2	-14.923704743376\\
2.5	-15.3743593396194\\
3	-15.7924915084934\\
3.5	-16.1825613539112\\
4	-16.5481897813199\\
};
\addlegendentry{TTL OPT$|_{\text{idealized}}$}


\addplot [color=mycolor1, mark=x, mark options={solid, mycolor1}]
  table[row sep=crcr]{%
1e-05	-15.3526132166498\\
0.5	-13.9515298303088\\
1	-13.9252715090326\\
1.5	-13.9652532706859\\
2	-14.0821991042476\\
2.5	-14.1551023385244\\
3	-14.2520764103626\\
3.5	-14.289055600185\\
4	-14.375845955716\\
};
\addlegendentry{LRU}

\addplot [color=mycolor2,mark size=1.5pt, dashdotted, mark=asterisk, mark options={solid, mycolor2}]
  table[row sep=crcr]{%
1e-05	-14.6681302697846\\
0.5	-14.0052630793324\\
1	-14.0808600149119\\
1.5	-14.2092796595985\\
2	-14.182574265935\\
2.5	-14.4224008842581\\
3	-14.5228536890978\\
3.5	-14.4816952602745\\
4	-14.6307737902499\\
};
\addlegendentry{Random}

\addplot [color=mycolor3, mark size=1.5pt,mark=square, mark options={solid, mycolor3}]
  table[row sep=crcr]{%
1e-05	-15.4273922282341\\
0.5	-14.082766033149\\
1	-14.1234705525138\\
1.5	-14.1429371580269\\
2	-14.2789708426048\\
2.5	-14.362604227184\\
3	-14.4839950665695\\
3.5	-14.5288936115315\\
4	-14.6460905815549\\
};
\addlegendentry{FIFO}

\end{axis}

\end{tikzpicture}%

%% file: Figs/tree.tex
\begin{tikzpicture}[sibling distance=0.5cm, level distance = 2.6cm]
  \tikzset{
    parent/.style={draw, rectangle, rounded corners=5pt, minimum height=0.9cm, minimum width=2cm},
    child/.style={draw, circle, minimum size=1.2cm}
  }
  
    \node [child,font=\Large] (parent1) {\rotatebox{90}{$C_{{n_l}+1}$}}
    child { node [child,font=\Large] (child1) {\rotatebox{90}{$C_1$}} } 
    child[dotted, shorten >= 0.3cm]
    child[dotted, shorten >= 0.3cm]
    child[dotted, shorten >= 0.3cm]
    child { node [child, font=\Large] (child2) {\rotatebox{90}{$C_{n_l}$}} }; 
    
  \node [above=1.5cm of parent1, parent,font=\Large] (parent2) {Server};
  
  \draw (parent1) -- (parent2);
  
  \node [below=0.7cm of child1,font=\Large] (word1) {\rotatebox{90}{$\lambda_{I,i1}$}};
  \node [below=0.7cm of child2,font=\Large] (word2) {\rotatebox{90}{$\lambda_{I,in_l}$}};
    
  \node [left=0.4cm of parent1, xshift=0.55cm, yshift=-1.3cm,font=\Large] (word3) {\rotatebox{90}{$d_1$}};
  \node [right=0.4cm of parent1,xshift=-0.55cm, yshift=-1.3cm,font=\Large] (word3) {\rotatebox{90}{$d_{n_l}$}};
  \node [above=0.09cm of parent1, yshift=0.1cm,xshift=0.3cm,font=\Large] (word3) {\rotatebox{90}{$d_{n_l+1}$}};

  \draw[-{Latex[length=2mm,width=2mm]}, line width=0.5pt] (word1) -- (child1);
  \draw[-{Latex[length=2mm,width=2mm]}, line width=0.5pt] (word2) -- (child2);

\end{tikzpicture}

%% file: Figs/MAP_single_1.tex
\ifx\XFigwidth\undefined\dimen1=0pt\else\dimen1\XFigwidth\fi
\divide\dimen1 by 1831
\ifx\XFigheight\undefined\dimen3=0pt\else\dimen3\XFigheight\fi
\divide\dimen3 by 1348
\ifdim\dimen1=0pt\ifdim\dimen3=0pt\dimen1=4143sp\dimen3\dimen1
  \else\dimen1\dimen3\fi\else\ifdim\dimen3=0pt\dimen3\dimen1\fi\fi
\tikzpicture[x=+\dimen1, y=+\dimen3]
{\ifx\XFigu\undefined\catcode`\@11
\def\temp{\alloc@1\dimen\dimendef\insc@unt}\temp\XFigu\catcode`\@12\fi}
\XFigu4143sp
\ifdim\XFigu<0pt\XFigu-\XFigu\fi
\pgfdeclarearrow{
  name = xfiga0,
  parameters = {
    \the\pgfarrowlinewidth \the\pgfarrowlength \the\pgfarrowwidth},
  defaults = {
	  line width=+7.5\XFigu, length=+120\XFigu, width=+60\XFigu},
  setup code = {
    \dimen7 2.15\pgfarrowlength\pgfmathveclen{\the\dimen7}{\the\pgfarrowwidth}
    \dimen7 2\pgfarrowwidth\pgfmathdivide{\pgfmathresult}{\the\dimen7}
    \dimen7 \pgfmathresult\pgfarrowlinewidth
    \pgfarrowssettipend{+\dimen7}
    \pgfarrowssetbackend{+-\pgfarrowlength}
    \dimen9 -0.5\pgfarrowlinewidth
    \pgfarrowssetvisualbackend{+\dimen9}
    \pgfarrowssetlineend{+-0.5\pgfarrowlinewidth}
    \pgfarrowshullpoint{+\dimen7}{+0pt}
    \pgfarrowsupperhullpoint{+-\pgfarrowlength}{+0.5\pgfarrowwidth}
    \pgfarrowssavethe\pgfarrowlinewidth
    \pgfarrowssavethe\pgfarrowlength
    \pgfarrowssavethe\pgfarrowwidth
  },
  drawing code = {\pgfsetdash{}{+0pt}
    \ifdim\pgfarrowlinewidth=\pgflinewidth\else\pgfsetlinewidth{+\pgfarrowlinewidth}\fi
    \pgfpathmoveto{\pgfqpoint{-\pgfarrowlength}{0.5\pgfarrowwidth}}
    \pgfpathlineto{\pgfqpoint{0pt}{0pt}}
    \pgfpathlineto{\pgfqpoint{-\pgfarrowlength}{-0.5\pgfarrowwidth}}
    \pgfusepathqstroke
  }
}
\pgfdeclarearrow{
  name = xfiga1,
  parameters = {
    \the\pgfarrowlinewidth \the\pgfarrowlength \the\pgfarrowwidth},
  defaults = {
	  line width=+7.5\XFigu, length=+120\XFigu, width=+60\XFigu},
  setup code = {
    \dimen7 2.15\pgfarrowlength\pgfmathveclen{\the\dimen7}{\the\pgfarrowwidth}
    \dimen7 2\pgfarrowwidth\pgfmathdivide{\pgfmathresult}{\the\dimen7}
    \dimen7 \pgfmathresult\pgfarrowlinewidth
    \pgfarrowssettipend{+\dimen7}
    \pgfarrowssetbackend{+-\pgfarrowlength}
    \dimen9 -0.5\pgfarrowlinewidth
    \pgfarrowssetvisualbackend{+\dimen9}
    \pgfarrowssetlineend{+-0.5\pgfarrowlinewidth}
    \pgfarrowshullpoint{+\dimen7}{+0pt}
    \pgfarrowsupperhullpoint{+-\pgfarrowlength}{+0.5\pgfarrowwidth}
    \pgfarrowssavethe\pgfarrowlinewidth
    \pgfarrowssavethe\pgfarrowlength
    \pgfarrowssavethe\pgfarrowwidth
  },
  drawing code = {\pgfsetdash{}{+0pt}
        \pgftransformrotate{18}
\ifdim\pgfarrowlinewidth=\pgflinewidth\else\pgfsetlinewidth{+\pgfarrowlinewidth}\fi
    \pgfpathmoveto{\pgfqpoint{-\pgfarrowlength}{0.5\pgfarrowwidth}}
    \pgfpathlineto{\pgfqpoint{0pt}{0pt}}
    \pgfpathlineto{\pgfqpoint{-\pgfarrowlength}{-0.5\pgfarrowwidth}}
    \pgfusepathqstroke
  }
}

\tikzset{inner sep=+0pt, outer sep=+0pt}
\pgfsetlinewidth{+7.5\XFigu}
\draw  (540,-1440) circle [radius=+262];
\draw  (1718,-1440) circle [radius=+262];
\draw  (1133,-2250) circle [radius=+262];
\pgfsetarrows{[line width=15\XFigu, width=75\XFigu]}
\pgfsetarrowsend{xfiga0}
\draw[dashed] (810,-1440)--(1440,-1440);
\draw (1645,-1685)--(1305,-2070);
\draw (900,-2115)--(585,-1710);
\pgfsetarrowsend{xfiga1}
\draw [dashed](1718,-1178) arc [start angle=160, end angle=-87, radius=240];

\pgftext[base,left,at=\pgfqpointxy{500}{-1495}] {\fontsize{12}{14.4}$0$}
\pgftext[base,left,at=\pgfqpointxy{1645}{-1495}] {\fontsize{12}{14.4}$F$}
\pgftext[base,left,at=\pgfqpointxy{1085}{-2305}] {\fontsize{12}{14.4}$1$}
\pgftext[base,left,at=\pgfqpointxy{1035}{-1350}] {\fontsize{12}{14.4}$\lambda$}
\pgftext[base,left,at=\pgfqpointxy{1940}{-1250}] {\fontsize{12}{14.4}$\lambda$}
\pgftext[base,left,at=\pgfqpointxy{1290}{-1790}] {\fontsize{12}{14.4}$\mu_F$}
\pgftext[base,left,at=\pgfqpointxy{740}{-1800}] {\fontsize{12}{14.4}$\mu$}
\endtikzpicture%

%% file: Figs/MAP_Single_2.tex
\ifx\XFigwidth\undefined\dimen1=0pt\else\dimen1\XFigwidth\fi
\divide\dimen1 by 3901
\ifx\XFigheight\undefined\dimen3=0pt\else\dimen3\XFigheight\fi
\divide\dimen3 by 3270
\ifdim\dimen1=0pt\ifdim\dimen3=0pt\dimen1=4143sp\dimen3\dimen1
  \else\dimen1\dimen3\fi\else\ifdim\dimen3=0pt\dimen3\dimen1\fi\fi
\tikzpicture[x=+\dimen1, y=+\dimen3]
{\ifx\XFigu\undefined\catcode`\@11
\def\temp{\alloc@1\dimen\dimendef\insc@unt}\temp\XFigu\catcode`\@12\fi}
\XFigu4143sp
\ifdim\XFigu<0pt\XFigu-\XFigu\fi
\pgfdeclarearrow{
  name = xfiga0,
  parameters = {
    \the\pgfarrowlinewidth \the\pgfarrowlength \the\pgfarrowwidth},
  defaults = {
	  line width=+7.5\XFigu, length=+120\XFigu, width=+60\XFigu},
  setup code = {
    \dimen7 2.15\pgfarrowlength\pgfmathveclen{\the\dimen7}{\the\pgfarrowwidth}
    \dimen7 2\pgfarrowwidth\pgfmathdivide{\pgfmathresult}{\the\dimen7}
    \dimen7 \pgfmathresult\pgfarrowlinewidth
    \pgfarrowssettipend{+\dimen7}
    \pgfarrowssetbackend{+-\pgfarrowlength}
    \dimen9 -0.5\pgfarrowlinewidth
    \pgfarrowssetvisualbackend{+\dimen9}
    \pgfarrowssetlineend{+-0.5\pgfarrowlinewidth}
    \pgfarrowshullpoint{+\dimen7}{+0pt}
    \pgfarrowsupperhullpoint{+-\pgfarrowlength}{+0.5\pgfarrowwidth}
    \pgfarrowssavethe\pgfarrowlinewidth
    \pgfarrowssavethe\pgfarrowlength
    \pgfarrowssavethe\pgfarrowwidth
  },
  drawing code = {\pgfsetdash{}{+0pt}
    \ifdim\pgfarrowlinewidth=\pgflinewidth\else\pgfsetlinewidth{+\pgfarrowlinewidth}\fi
    \pgfpathmoveto{\pgfqpoint{-\pgfarrowlength}{0.5\pgfarrowwidth}}
    \pgfpathlineto{\pgfqpoint{0pt}{0pt}}
    \pgfpathlineto{\pgfqpoint{-\pgfarrowlength}{-0.5\pgfarrowwidth}}
    \pgfusepathqstroke
  }
}
\tikzset{inner sep=+0pt, outer sep=+0pt}
\pgfsetarrows{[line width=15\XFigu, width=75\XFigu]}
\pgfsetarrowsend{xfiga0}
\pgfsetlinewidth{+7.5\XFigu}
\draw (3560,-3320) arc[start angle=+-67.78, end angle=+-114, radius=+3071.6];
\draw (3560,-1215) arc[start angle=+67.78, end angle=+114, radius=+3071.6];
\draw (2625,-2940) [dashed]arc[start angle=+-39, end angle=+38, radius=+1015.9];
\draw (3900,-2940) arc[start angle=+-39, end angle=+38, radius=+1015.9];
\draw  (1035,-1620) circle [radius=+400];
\draw  (1035,-2912) circle [radius=+400];
\draw  (2233,-1620) circle [radius=+400];
\draw  (2233,-2912) circle [radius=+400];
\draw  (3493,-1620) circle [radius=+400];
\draw  (3493,-2912) circle [radius=+400];
\draw (1035,-2025)--(1035,-2475);
\draw [dashed](1350,-2655)--(1980,-1935);
\draw (2625,-1665)--(3085,-1665);
\draw (2625,-2970)--(3085,-2970);
\draw (2250,-2025)--(2250,-2475);
\draw (3465,-2025)--(3465,-2475);
\pgftext[base,left,at=\pgfqpointxy{960}{-1700}]{\fontsize{16}{18.4}$0_a$}
\pgftext[base,left,at=\pgfqpointxy{960}{-2995}]{\fontsize{16}{18.4}$0_B$}
\pgftext[base,left,at=\pgfqpointxy{2100}{-1700}] {\fontsize{16}{18.4}$F_a$}
\pgftext[base,left,at=\pgfqpointxy{2100}{-2995}] {\fontsize{16}{18.4}$F_b$}
\pgftext[base,left,at=\pgfqpointxy{3420}{-1700}] {\fontsize{16}{18.4}$1_a$}
\pgftext[base,left,at=\pgfqpointxy{3420}{-2995}] {\fontsize{16}{18.4}$1_b$}
\pgftext[base,left,at=\pgfqpointxy{815}{-2305}] {\fontsize{16}{18.4}$\lambda$}
\pgftext[base,left,at=\pgfqpointxy{2205}{-870}] {\fontsize{16}{18.4}$\mu_T$}
\pgftext[base,left,at=\pgfqpointxy{2730}{-1540}] {\fontsize{16}{18.4}$\mu_F$}
\pgftext[base,left,at=\pgfqpointxy{4175}{-2305}] {\fontsize{16}{18.4}$\lambda$}
\pgftext[base,left,at=\pgfqpointxy{3520}{-2305}] {\fontsize{16}{18.4}$\lambda$}
\pgftext[base,left,at=\pgfqpointxy{2730}{-3150}] {\fontsize{16}{18.4}$\mu_F$}
\pgftext[base,left,at=\pgfqpointxy{2205}{-3700}] {\fontsize{16}{18.4}$\mu_T$}
\pgftext[base,left,at=\pgfqpointxy{2670}{-2305}] {\fontsize{16}{18.4}$\lambda$}
\pgftext[base,left,at=\pgfqpointxy{2030}{-2305}] {\fontsize{16}{18.4}$\lambda$}
\pgftext[base,left,at=\pgfqpointxy{1485}{-2305}] {\fontsize{16}{18.4}$\lambda$}
\endtikzpicture%

%% file: Evaluations/Delay_Impact_single_cache/Throughput_Occ_errors_arxiv.tex
%
%
\definecolor{mycolor1}{rgb}{0.00000,0.44700,0.74100}%
\definecolor{mycolor2}{rgb}{0.85000,0.32500,0.09800}%
\begin{tikzpicture}

\begin{axis}[%
width=2in,
height=1in,
at={(0.772in,0.473in)},
scale only axis,
xmin=0,
xmax=4,
xlabel style={font=\color{white!15!black}},
xlabel={delay ratio $\rho_d$},
separate axis lines,
ymin=0,
ymax=30,
ylabel style={align=center},
ylabel style={font=\color{white!15!black}},
ylabel style={align=center},
ylabel={Offloading $[\%]$ \\ improvement },
ylabel style={yshift=-0.3 cm},
xlabel style={yshift=0.15cm},
axis background/.style={fill=white},
legend style={nodes={scale=0.7}},
legend style={at={(0.01,0.83)},anchor=west},
xmajorgrids,
ymajorgrids
]
\draw[line width=0.005 \linewidth](current axis.south west)rectangle(current axis.north east);

\addplot [color=mycolor2, mark=o]
  table[row sep=crcr]{%
0	0\\
0.5	4.03000511016322\\
1	7.50581983219101\\
1.5	10.5384734517489\\
2	13.2106830587722\\
2.5	15.5853995930138\\
3	17.711447006655\\
3.5	19.6273198282521\\
4	21.3638086609854\\
};

\end{axis}

\end{tikzpicture}%

%% file: Evaluations/Delay_Impact_single_cache/TTL_values_arxiv.tex
%
%
\definecolor{mycolor1}{rgb}{0.00000,0.44700,0.74100}%
\definecolor{mycolor2}{rgb}{0.85000,0.32500,0.09800}%
\definecolor{mycolor3}{rgb}{0.92900,0.69400,0.12500}%
\definecolor{mycolor4}{rgb}{0.49400,0.18400,0.55600}%
\definecolor{mycolor5}{rgb}{0.46600,0.67400,0.18800}%
\definecolor{mycolor6}{rgb}{0.30100,0.74500,0.93300}%
\definecolor{mycolor7}{rgb}{0.63500,0.07800,0.18400}%
\begin{tikzpicture}

\begin{loglogaxis}[%
width=2in,
height=1in,
scale only axis,
xmin=1,
xmax=100,
xlabel style={font=\color{white!15!black}},
xlabel={obj ID},
ymin=1,
ymax=20,
ylabel style={font=\color{white!15!black}},
ylabel={Optimal TTL [s]},
ylabel style={yshift=-0.2 cm},
axis background/.style={fill=white},
axis x line*=bottom,
axis y line*=left,
xlabel style={yshift=0.15cm},
xmajorgrids,
ymajorgrids,
xminorgrids,
legend style={legend cell align=left, align=left, draw=white!15!black}
]
\draw[line width=0.005 \linewidth](current axis.south west)rectangle(current axis.north east);

\addplot [color=mycolor1, only marks, mark=asterisk, mark options={scale=0.5,solid, mycolor1}]
  table[row sep=crcr]{%
1   50\\
2	4.57950190435304\\
3	2.64922045369728\\
4	2.16059137877953\\
5	1.93510657072846\\
6	1.80434215146329\\
7	1.71854131678233\\
8	1.65768999831488\\
9	1.61215831454346\\
10	1.57672868304094\\
11	1.54832341950052\\
12	1.52500706513346\\
13	1.50550000143384\\
14	1.48892147678736\\
15	1.47464488940731\\
16	1.46221193772637\\
17	1.45127944194676\\
18	1.44158517722411\\
19	1.43292522347984\\
20	1.42513853685623\\
21	1.41809618768499\\
22	1.41169369445312\\
23	1.40584546030901\\
24	1.40048066745656\\
25	1.39554020148985\\
26	1.3909743157023\\
27	1.38674083522758\\
28	1.38280376052477\\
29	1.37913217007038\\
30	1.37569934986927\\
31	1.37248209677521\\
32	1.36946015633642\\
33	1.36661576572885\\
34	1.36393327949171\\
35	1.3613988610313\\
36	1.35900022675745\\
37	1.35672643263734\\
38	1.35456769516041\\
39	1.3525152403935\\
40	1.3505611761006\\
41	1.34869838290581\\
42	1.34692042126137\\
43	1.34522145159868\\
44	1.34359616552634\\
45	1.34203972632706\\
46	1.34054771731506\\
47	1.33911609686477\\
48	1.33774115912339\\
49	1.33641949958405\\
50	1.3351479848297\\
51	1.33392372586822\\
52	1.33274405456935\\
53	1.3316065027892\\
54	1.33050878383017\\
55	1.32944877593605\\
56	1.32842450756543\\
57	1.3274341442232\\
58	1.32647597666036\\
59	1.32554841027855\\
60	1.32464995559786\\
61	1.32377921966484\\
62	1.32293489829408\\
63	1.32211576905005\\
64	1.32132068488792\\
65	1.32054856838188\\
66	1.31979840647854\\
67	1.31906924572022\\
68	1.31836018788956\\
69	1.31767038603264\\
70	1.31699904082259\\
71	1.31634539723003\\
72	1.31570874147051\\
73	1.3150883982023\\
74	1.31448372795072\\
75	1.31389412473802\\
76	1.3133190138997\\
77	1.3127578500705\\
78	1.31221011532449\\
79	1.31167531745606\\
80	1.31115298838911\\
81	1.31064268270357\\
82	1.31014397626917\\
83	1.30965646497744\\
84	1.30917976356376\\
85	1.30871350451208\\
86	1.3082573370356\\
87	1.30781092612724\\
88	1.30737395167445\\
89	1.30694610763332\\
90	1.30652710125721\\
91	1.30611665237605\\
92	1.30571449272214\\
93	1.30532036529919\\
94	1.30493402379121\\
95	1.30455523200849\\
96	1.30418376336795\\
97	1.30381940040524\\
98	1.30346193431646\\
99	1.30311116452757\\
100	1.3027668982891\\
};
\addlegendentry{$\rho_d$=0}

\addplot [color=mycolor5,only marks, mark=o, mark options={scale=0.5,solid, mycolor5}]
  table[row sep=crcr]{%
1   50\\
2	9.83991161376679\\
3	4.84934421782971\\
4	3.58603824374082\\
5	3.00306784713742\\
6	2.66498839729259\\
7	2.44315820214689\\
8	2.28583270894302\\
9	2.16811473549264\\
10	2.07651469744135\\
11	2.00307552227428\\
12	1.9427932575793\\
13	1.89235956628526\\
14	1.84949734534889\\
15	1.81258657295991\\
16	1.78044235219538\\
17	1.75217742297983\\
18	1.72711382689368\\
19	1.70472434717618\\
20	1.68459261962782\\
21	1.66638530835844\\
22	1.64983228595457\\
23	1.63471224957274\\
24	1.62084210626971\\
25	1.60806902114996\\
26	1.5962643786502\\
27	1.58531913950907\\
28	1.57514023020557\\
29	1.56564770596924\\
30	1.55677250021008\\
31	1.54845462331738\\
32	1.5406417092606\\
33	1.5332878338852\\
34	1.52635254728574\\
35	1.51980007621746\\
36	1.51359866258433\\
37	1.50772001159323\\
38	1.50213882887536\\
39	1.49683243023195\\
40	1.49178041101153\\
41	1.48696436472164\\
42	1.48236764250344\\
43	1.47797514668915\\
44	1.47377315292103\\
45	1.46974915631161\\
46	1.46589173792637\\
47	1.46219044851458\\
48	1.45863570693528\\
49	1.45521871114972\\
50	1.45193135999715\\
51	1.44876618425489\\
52	1.44571628571847\\
53	1.44277528322975\\
54	1.43993726474336\\
55	1.4371967446547\\
56	1.43454862572582\\
57	1.43198816503935\\
58	1.42951094349016\\
59	1.42711283839192\\
60	1.42478999883231\\
61	1.42253882345934\\
62	1.42035594042259\\
63	1.41823818922846\\
64	1.416182604299\\
65	1.41418640004992\\
66	1.41224695732587\\
67	1.41036181105065\\
68	1.40852863896689\\
69	1.40674525135406\\
70	1.40500958162699\\
71	1.40331967772759\\
72	1.4016736942328\\
73	1.40006988510953\\
74	1.39850659705573\\
75	1.39698226337254\\
76	1.39549539831867\\
77	1.39404459190314\\
78	1.39262850507699\\
79	1.39124586528873\\
80	1.38989546237144\\
81	1.38857614473332\\
82	1.38728681582532\\
83	1.38602643086294\\
84	1.38479399378069\\
85	1.38358855440039\\
86	1.38240920579576\\
87	1.38125508183751\\
88	1.38012535490481\\
89	1.37901923374991\\
90	1.37793596150404\\
91	1.37687481381394\\
92	1.37583509709878\\
93	1.37481614691891\\
94	1.37381732644765\\
95	1.37283802503888\\
96	1.37187765688346\\
97	1.37093565974791\\
98	1.37001149378961\\
99	1.36910464044332\\
100	1.36821460137363\\
};
\addlegendentry{$\rho_d$=2}

\addplot [color=mycolor2, mark=square, only marks, mark options={scale=0.5,solid, mycolor2}]
  table[row sep=crcr]{%
1   50\\
2	15.1003475405011\\
3	7.04947896210373\\
4	5.01149222512278\\
5	4.07103445609926\\
6	3.5256389409041\\
7	3.16777870622705\\
8	2.91397855663218\\
9	2.72407393308026\\
10	2.57630320800815\\
11	2.45782989633939\\
12	2.36058153672194\\
13	2.27922106339229\\
14	2.21007501490735\\
15	2.15052994447366\\
16	2.09867435618504\\
17	2.0530769069718\\
18	2.01264390276344\\
19	1.97652482850275\\
20	1.94404799837356\\
21	1.91467566924316\\
22	1.88797206697048\\
23	1.8635801820425\\
24	1.84120464580819\\
25	1.82059890241448\\
26	1.80155546704742\\
27	1.78389843571677\\
28	1.7674776606364\\
29	1.75216417354408\\
30	1.73784655504345\\
31	1.72442802887557\\
32	1.71182411727082\\
33	1.69996073460346\\
34	1.68877262639957\\
35	1.6782020826538\\
36	1.66819787066692\\
37	1.65871434479892\\
38	1.6497106997453\\
39	1.64115034097217\\
40	1.63300035135009\\
41	1.62523103721373\\
42	1.61781554034213\\
43	1.61072950492212\\
44	1.60395079058756\\
45	1.59745922424254\\
46	1.59123638466881\\
47	1.58526541495843\\
48	1.57953085865299\\
49	1.57401851615494\\
50	1.56871531853489\\
51	1.56360921631688\\
52	1.55868908120096\\
53	1.55394461899529\\
54	1.54936629228857\\
55	1.54494525161105\\
56	1.54067327401262\\
57	1.53654270813908\\
58	1.53254642501569\\
59	1.5286777738555\\
60	1.52493054230198\\
61	1.52129892059358\\
62	1.51777746920452\\
63	1.51436108957347\\
64	1.5110449975803\\
65	1.50782469947368\\
66	1.50469596998825\\
67	1.50165483242179\\
68	1.49869754046975\\
69	1.49582056163836\\
70	1.49302056207773\\
71	1.49029439269521\\
72	1.48763907642335\\
73	1.48505179653242\\
74	1.48252988588791\\
75	1.48007081706505\\
76	1.4776721932412\\
77	1.47533173979538\\
78	1.47304729655151\\
79	1.47081681060824\\
80	1.46863832970421\\
81	1.46650999607228\\
82	1.46443004074135\\
83	1.46239677824762\\
84	1.46040860172172\\
85	1.45846397832043\\
86	1.45656144497516\\
87	1.45469960443182\\
88	1.45287712155873\\
89	1.4510927199019\\
90	1.4493451784681\\
91	1.44763332871863\\
92	1.44595605175745\\
93	1.44431227569952\\
94	1.44270097320546\\
95	1.44112115917094\\
96	1.43957188855893\\
97	1.43805225436509\\
98	1.43656138570643\\
99	1.43509844602493\\
100	1.43366263139775\\
};
\addlegendentry{$\rho_d$=4}

\end{loglogaxis}

\end{tikzpicture}%

%% file: Evaluations/Delay_Impact_on_three_caches/Hierarchy_vs_individual_arxiv.tex
%
%
\definecolor{mycolor1}{rgb}{0.00000,0.44700,0.74100}%
\definecolor{mycolor2}{rgb}{0.85000,0.32500,0.09800}%
\begin{tikzpicture}

\begin{axis}[%
width=2 in,
height=1.4in,
scale only axis,
xmin=0,
xmax=4,
xlabel style={font=\color{white!15!black}},
xlabel={delay ratio $\rho_d$},
ymin=-19,
ymax=-12,
ylabel style={font=\color{white!15!black}},
ylabel={Aggregate utility},
axis background/.style={fill=white},
ylabel style={yshift=-0.3cm},
xlabel style={yshift=0.15cm},
axis x line*=bottom,
axis y line*=left,
xmajorgrids,
ymajorgrids,
legend style={legend cell align=left, align=left, draw=white!15!black,},
legend style={nodes={scale=0.7}},
legend pos=south west,
]
\draw[line width=0.005 \linewidth](current axis.south west)rectangle(current axis.north east);

\addplot [color=mycolor2, mark=o, mark options={solid, mycolor2}]
  table[row sep=crcr]{%
1e-05	-12.7017615574456\\
0.5	-12.6651789118222\\
1	-12.6627053022838\\
1.5	-12.6823143223598\\
2	-12.6172116485529\\
2.5	-12.6563563498063\\
3	-12.6572982550923\\
3.5	-12.6081642222134\\
4	-12.5920845215303\\
};
\addlegendentry{TTL OPT$|_{\text{delay}}$}

\addplot [color=mycolor2, mark=*]
  table[row sep=crcr]{%
1e-05	-12.7017615574456\\
0.5	-13.3187957438792\\
1	-13.9019533961636\\
1.5	-14.4350490866435\\
2	-14.923704743376\\
2.5	-15.3743593396194\\
3	-15.7924915084934\\
3.5	-16.1825613539112\\
4	-16.5481897813199\\
};
\addlegendentry{TTL OPT$|_{\text{idealize}}$}


\addplot [color=mycolor1, mark=x, mark options={solid, mycolor1}]
  table[row sep=crcr]{%
1e-05	-15.3526132166498\\
0.5	-13.9515298303088\\
1	-13.9252715090326\\
1.5	-13.9652532706859\\
2	-14.0821991042476\\
2.5	-14.1551023385244\\
3	-14.2520764103626\\
3.5	-14.289055600185\\
4	-14.375845955716\\
};
\addlegendentry{LRU}
\end{axis}

\end{tikzpicture}%

%% file: Evaluations/Delay_Impact_on_three_caches/Error_EKAD_arxiv.tex
%
%
\definecolor{mycolor1}{rgb}{0.00000,0.44700,0.74100}%
\definecolor{mycolor2}{rgb}{0.85000,0.32500,0.09800}%
\definecolor{mycolor3}{rgb}{0.92900,0.69400,0.12500}%

\begin{tikzpicture}

\begin{axis}[%
width=2in,
height=1.4in,
scale only axis,
xmin=0,
xmax=4,
ymin=-0.5,
ymax=15,
xlabel style={font=\color{white!15!black}},
xlabel={delay ratio $\rho_d$},
ylabel style={yshift=-0.55cm},
ylabel style={font=\color{white!15!black}},
xlabel style={yshift=0.15cm},
ylabel={Utility loss $[\%]$},
axis background/.style={fill=white},
axis x line*=bottom,
axis y line*=left,
xmajorgrids,
ymajorgrids,
legend style={legend cell align=left, align=left, draw=white!15!black},
legend style={nodes={scale=0.8}},
]
\draw[line width=0.005 \linewidth](current axis.south west)rectangle(current axis.north east);

\addplot [color=mycolor1, mark=asterisk, mark options={solid, mycolor1}]
  table[row sep=crcr]{%
1e-05	0.105569895977728\\
0.5	0.174821071933321\\
1	0.395600271080326\\
1.5	0.282360470866945\\
2	0.0392234593939859\\
2.5	0.273704653307235\\
3	0.214910874440311\\
3.5	0.364892427710775\\
4	0.597501837158951\\
};
\addlegendentry{TTL$_{\min,\text{extnd}}$}

\addplot [color=mycolor2, mark=o, mark options={solid, mycolor2}]
  table[row sep=crcr]{%
1e-05	9.2909342816717\\
0.5	9.06146705708189\\
1	9.13321960498108\\
1.5	8.61886206588493\\
2	8.88473247711281\\
2.5	9.27710007237788\\
3	9.19725176548575\\
3.5	9.20323936772281\\
4	9.25971906846603\\
};
\addlegendentry{TTL$_{\min}$}


\end{axis}

\end{tikzpicture}%

%% file: Evaluations/Run_time.tex
%
%
\definecolor{mycolor1}{rgb}{0.00000,0.44700,0.74100}%
\definecolor{mycolor2}{rgb}{0.85000,0.32500,0.09800}%
\begin{tikzpicture}

\begin{axis}[%
width=2in,
height=0.8in,
scale only axis,
scale only axis,
xmin=1,
xmax=20,
xlabel style={font=\color{white!15!black}},
ymode=log,
ymin=0,
ymax=1000,
yminorticks=true,
ylabel style={font=\color{white!15!black}},
 ylabel style={align=center},
 ylabel={Execution \\ time (s)},
axis background/.style={fill=white},
xmajorgrids,
ymajorgrids,
xticklabels={},
]
\draw[line width=0.002 \linewidth](current axis.south west)rectangle(current axis.north east);
\addplot [color=mycolor1,only marks,mark=asterisk,mark options={solid, thick, mycolor1}]
  table[row sep=crcr]{%
1	0.03014\\
2	0.20828\\
3	3.16\\
4	390.88\\
};
\addplot [color=mycolor2,only marks,mark=o,mark options={solid, thick, mycolor2}]
  table[row sep=crcr]{%
1  0.1496429\\
2 0.144 \\
3 0.146 \\
4 0.147 \\
5 0.148 \\
6 0.151 \\
7 0.150 \\
8 0.155 \\
9 0.156 \\
10 0.165 \\
11 0.156 \\
12 0.156 \\
13 0.158 \\
13 0.161 \\
14 0.163 \\
15 0.168 \\
16 0.166 \\
17 0.168 \\
18 0.165 \\
19 0.168 \\
20 0.167 \\
};

\end{axis}

\end{tikzpicture}%

%% file: Evaluations/Acc.tex
%
%
\definecolor{mycolor1}{rgb}{0.00000,0.44700,0.74100}%
\definecolor{mycolor2}{rgb}{0.85000,0.32500,0.09800}%
\begin{tikzpicture}

\begin{axis}[%
width=2in,
height=0.7in,
scale only axis,
xmin=1,
xmax=6,
xlabel style={font=\color{white!15!black}},
xlabel={Number of caches},
ylabel style={font=\color{white!15!black}},
ylabel={Aggregate \\ Utility},
 ylabel style={align=center},
 xlabel style={yshift=0.2cm},
axis background/.style={fill=white},
title style={font=\bfseries},
axis x line*=bottom,
axis y line*=left,
 xmajorgrids,
ymajorgrids,
legend style={legend cell align=left, align=left, draw=white!15!black}
]
\draw[line width=0.002 \linewidth](current axis.south west)rectangle(current axis.north east);
\addplot [color=mycolor1, mark=asterisk, only marks, mark options={thick,solid, mycolor1}]
  table[row sep=crcr]{%
1	-4.39929510151041\\
2	-2.74914522813561\\
3	-5.43982560043335\\
4	-8.27927022529321\\
};
\addlegendentry{$\text{TTL}_{\min}$}

\addplot [color=mycolor2, mark=o, only marks, mark options={solid, thick, mycolor2}]
  table[row sep=crcr]{%
1    -4.28964341 \\
2    -2.71243650 \\
3    -5.46657421 \\
4    -8.31910691 \\
5   -11.31662856 \\
6   -14.38662029 \\
};
\addlegendentry{TTL\textsubscript{GNN}}


\end{axis}

\end{tikzpicture}%

%% file: Evaluations/Trace/Trace_offloading.tex
%
%
\definecolor{mycolor1}{rgb}{0.00000,0.44700,0.74100}%
\definecolor{mycolor2}{rgb}{0.85000,0.32500,0.09800}%
\definecolor{mycolor3}{rgb}{0.92900,0.69400,0.12500}
\begin{tikzpicture}

\begin{axis}[%
width=1.3in,
height=0.6in,
scale only axis,
xmin=0,
xmax=4,
xlabel style={font=\color{white!15!black}},
ymin=0,
ymax=30,
ylabel style={align=center},
ylabel style={yshift=-0.3cm},
xlabel style={yshift=0.2cm},
ylabel style={font=\color{white!15!black}},
xlabel={delay ratio $\rho_d$},
ylabel={Offloading $[\%]$ \\ improvement },
axis background/.style={fill=white},
legend pos=outer north east,
xmajorgrids,
ymajorgrids
]
\addplot [color=mycolor1, mark=asterisk, mark options={solid, mycolor1}]
  table[row sep=crcr]{%
0	5.28244933160786\\
0.5	6.80907877169528\\
1	9.20988203872708\\
1.5	12.6252223574576\\
2	15.5426223895926\\
2.5	17.3986313961799\\
3	20.2675335749306\\
3.5	23.0895735116764\\
4	25.7134878525553\\
};
\addlegendentry{LRU}

\addplot [color=mycolor2,  mark=square, mark options={solid, mycolor2}]
  table[row sep=crcr]{%
0	5.4412913303823\\
0.5	7.19231435000834\\
1	9.64501004688549\\
1.5	12.7610644389172\\
2	15.7599606492864\\
2.5	19.1859866048426\\
3	22.2504498782681\\
3.5	24.5483002910649\\
4	27.9048850413178\\
};
\addlegendentry{FIFO}

\addplot [color=mycolor3, mark=o, mark options={solid, mycolor3}]
  table[row sep=crcr]{%
0	8.25739824698391\\
0.5	9.89700536610678\\
1	12.0830542531817\\
1.5	15.3001361055057\\
2	18.1997048696505\\
2.5	21.1076764554351\\
3	23.5736212554248\\
3.5	26.1134603767369\\
4	28.0566642307045\\
};
\addlegendentry{Random}

\end{axis}

\end{tikzpicture}%

%% file: main.bbl
\begin{thebibliography}{10}
\providecommand{\url}[1]{#1}
\csname url@samestyle\endcsname
\providecommand{\newblock}{\relax}
\providecommand{\bibinfo}[2]{#2}
\providecommand{\BIBentrySTDinterwordspacing}{\spaceskip=0pt\relax}
\providecommand{\BIBentryALTinterwordstretchfactor}{4}
\providecommand{\BIBentryALTinterwordspacing}{\spaceskip=\fontdimen2\font plus
\BIBentryALTinterwordstretchfactor\fontdimen3\font minus \fontdimen4\font\relax}
\providecommand{\BIBforeignlanguage}[2]{{%
\expandafter\ifx\csname l@#1\endcsname\relax
\typeout{** WARNING: IEEEtran.bst: No hyphenation pattern has been}%
\typeout{** loaded for the language `#1'. Using the pattern for}%
\typeout{** the default language instead.}%
\else
\language=\csname l@#1\endcsname
\fi
#2}}
\providecommand{\BIBdecl}{\relax}
\BIBdecl

\bibitem{Ramesh_Akamai_hierarchy}
B.~M. Maggs and R.~K. Sitaraman, ``Algorithmic nuggets in content delivery,'' \emph{ACM SIGCOMM Comput. Commun. Rev.}, vol.~45, no.~3, p. 52–66, 2015.

\bibitem{Dist_alg_CDN}
S.~Borst, V.~Gupta, and A.~Walid, ``Distributed caching algorithms for content distribution networks,'' in \emph{Proc. of IEEE INFOCOM}, 2010.

\bibitem{Deghan_utility}
M.~Dehghan, L.~Massoulié, D.~Towsley, D.~S. Menasché, and Y.~C. Tay, ``A utility optimization approach to network cache design,'' \emph{IEEE/ACM Transactions on Networking}, vol.~27, no.~3, pp. 1013--1027, 2019.

\bibitem{Traditional_caching_policies}
C.~Aggarwal, J.~Wolf, and P.~Yu, ``Caching on the world wide web,'' \emph{IEEE Transactions on Knowledge and Data Engineering}, vol.~11, no.~1, pp. 94--107, 1999.

\bibitem{Che:LRUtoTTL}
H.~Che, Y.~Tung, and Z.~Wang, ``Hierarchical web caching systems: modeling, design and experimental results,'' \emph{IEEE Journal on Selected Areas in Communications}, vol.~20, no.~7, pp. 1305--1314, 2002.

\bibitem{DeepCache_popularity_prediction}
A.~Narayanan, S.~Verma, E.~Ramadan, P.~Babaie, and Z.-L. Zhang, ``Deepcache: A deep learning based framework for content caching,'' in \emph{ACM Workshop on Network Meets AI \& ML - NetAI}, 2018, p. 48–53.

\bibitem{FNN_popularity_prediction}
V.~Fedchenko, G.~Neglia, and B.~Ribeiro, ``Feedforward neural networks for caching: N enough or too much?'' \emph{ACM SIGMETRICS Perform. Eval. Rev.}, vol.~46, no.~3, p. 139–142, 2019.

\bibitem{TTL_early_model}
J.~Jung, A.~Berger, and H.~Balakrishnan, ``Modeling {TTL}-based internet caches,'' in \emph{Proc. of IEEE INFOCOM}, 2003, pp. 417--426.

\bibitem{Cache_redund}
W.~K. Chai, D.~He, I.~Psaras, and G.~Pavlou, ``Cache “less for more” in information-centric networks,'' \emph{Computer Communications}, vol.~36, no.~7, pp. 758--770, 2013.

\bibitem{Berger:TTL_MAP}
D.~S. Berger, P.~Gland, S.~Singla, and F.~Ciucu, ``Exact analysis of {TTL} cache networks,'' \emph{Performance Evaluation}, vol.~79, pp. 2 -- 23, 2014.

\bibitem{Elsayed_cache_hierarchy_MAP}
K.~Elsayed and A.~Rizk, ``Time-to-live caching with network delays: Exact analysis and computable approximations,'' \emph{IEEE/ACM Transactions on Networking}, pp. 1--14, 2022.

\bibitem{FofackNNT14}
N.~C. Fofack, P.~Nain, G.~Neglia, and D.~Towsley, ``{Performance evaluation of hierarchical {TTL}-based cache networks},'' \emph{Computer Networks}, vol.~65, pp. 212--231, 2014.

\bibitem{Dehghan_cache_with_delay}
M.~Dehghan, B.~Jiang, A.~Dabirmoghaddam, and D.~Towsley, ``On the analysis of caches with pending interest tables,'' in \emph{Proc. of ACM Conference on Information-Centric Networking}, 2015, p. 69–78.

\bibitem{Don_hiera_ut}
N.~K. Panigrahy, J.~Li, F.~Zafari, D.~Towsley, and P.~Yu, ``A {TTL}-based approach for content placement in edge networks,'' in \emph{EAI International Conference on Performance Evaluation Methodologies and Tools}.\hskip 1em plus 0.5em minus 0.4em\relax Springer, 2021, pp. 1--21.

\bibitem{AtreSWB20}
N.~Atre, J.~Sherry, W.~Wang, and D.~S. Berger, ``Caching with delayed hits,'' in \emph{Proc. of {ACM} SIGCOMM}, 2020, pp. 495--513.

\bibitem{Kelly_Congestion}
F.~Kelly, ``Charging and rate control for elastic traffic,'' \emph{European Transactions on Telecommunications}, vol.~8, no.~1, pp. 33--37, 1997.

\bibitem{Srikant_control}
R.~Srikant and L.~Ying, \emph{Communication Networks: An Optimization, Control and Stochastic Networks Perspective}.\hskip 1em plus 0.5em minus 0.4em\relax Cambridge University Press, 2014.

\bibitem{Fofack:TTL}
N.~C. Fofack, P.~Nain, G.~Neglia, and D.~Towsley, ``Performance evaluation of hierarchical {TTL}-based cache networks,'' \emph{Computer Networks}, vol.~65, pp. 212 -- 231, 2014.

\bibitem{JiangNT18}
B.~Jiang, P.~Nain, and D.~Towsley, ``On the convergence of the {TTL} approximation for an {LRU} cache under independent stationary request processes,'' \emph{{ACM} Trans. Model. Perform. Evaluation Comput. Syst.}, vol.~3, no.~4, pp. 20:1--20:31, 2018.

\bibitem{Gelenbe73a}
E.~Gelenbe, ``A unified approach to the evaluation of a class of replacement algorithms,'' \emph{{IEEE} Trans. Computers}, vol.~22, no.~6, pp. 611--618, 1973.

\bibitem{NegliaCM18}
G.~Neglia, D.~Carra, and P.~Michiardi, ``Cache policies for linear utility maximization,'' \emph{{IEEE/ACM} Trans. Netw.}, vol.~26, no.~1, pp. 302--313, 2018.

\bibitem{TTL_heavy_tailet_opt}
A.~Ferragut, I.~Rodriguez, and F.~Paganini, ``Optimizing {TTL} caches under heavy-tailed demands,'' \emph{SIGMETRICS Perform. Eval. Rev.}, vol.~44, no.~1, p. 101–112, 2016.

\bibitem{Towsley_rate_vs_prob}
N.~K. Panigrahy, J.~Li, and D.~Towsley, ``Hit rate vs. hit probability based cache utility maximization,'' \emph{SIGMETRICS Perform. Eval. Rev.}, vol.~45, no.~2, p. 21–23, 2017.

\bibitem{Towsley_rate_vs_prob_Hazard}
N.~K. Panigrahy, J.~Li, D.~Towsley, and C.~Hollot, ``Network cache design under stationary requests: Exact analysis and poisson approximation,'' \emph{Computer Networks}, vol. 180, p. 107379, 2020.

\bibitem{Caching_hierarchy:Web_caching}
H.~Che, Z.~Wang, and Y.~Tung, ``Analysis and design of hierarchical web caching systems,'' in \emph{Proc. of IEEE INFOCOM 2001}, vol.~3, 2001, pp. 1416--1424.

\bibitem{Caching_hierarchy:CRAN_cooperative_caching}
T.~X. Tran and D.~Pompili, ``Octopus: A cooperative hierarchical caching strategy for cloud radio access networks,'' in \emph{Proc. of IEEE MASS}, 2016, pp. 154--162.

\bibitem{Caching_hierarchy:CDN}
S.~Saroiu, K.~P. Gummadi, R.~J. Dunn, S.~D. Gribble, and H.~M. Levy, ``An analysis of internet content delivery systems,'' \emph{ACM SIGOPS Oper. Syst. Rev.}, vol.~36, p. 315–327, 2003.

\bibitem{Caching_hierarchy:IPTV_video_cahcing}
L.~Chen, M.~Meo, and A.~Scicchitano, ``Caching video contents in {IPTV} systems with hierarchical architecture,'' in \emph{Proc. of IEEE International Conference on Communications}, 2009, pp. 1--6.

\bibitem{Caching_hierarchy:IPTV}
J.~Dai, Z.~Hu, B.~Li, J.~Liu, and B.~Li, ``Collaborative hierarchical caching with dynamic request routing for massive content distribution,'' in \emph{Proc. of IEEE INFOCOM}, 2012, pp. 2444--2452.

\bibitem{RL_cache}
V.~Kirilin, A.~Sundarrajan, S.~Gorinsky, and R.~K. Sitaraman, ``{RL-Cache}: Learning-based cache admission for content delivery,'' \emph{IEEE Journal on Selected Areas in Communications}, vol.~38, no.~10, pp. 2372--2385, 2020.

\bibitem{Zhong2018deep}
C.~Zhong, M.~C. Gursoy, and S.~Velipasalar, ``A deep reinforcement learning-based framework for content caching,'' in \emph{Proc. of IEEE Conference on Information Sciences and Systems (CISS)}, 2018, pp. 1--6.

\bibitem{Sadeghi2017optimal}
A.~Sadeghi, F.~Sheikholeslami, and G.~B. Giannakis, ``Optimal and scalable caching for 5g using reinforcement learning of space-time popularities,'' \emph{IEEE Journal of Selected Topics in Signal Processing}, vol.~12, no.~1, pp. 180--190, 2017.

\bibitem{Pedersen2021dynamic}
J.~Pedersen, F.~Br{\"a}nnstr{\"o}m, E.~Rosnes \emph{et~al.}, ``Dynamic coded caching in wireless networks using multi-agent reinforcement learning,'' \emph{arXiv preprint arXiv:2104.06724}, 2021.

\bibitem{Utility_DRL}
C.~Cho, S.~Shin, H.~Jeon, and S.~Yoon, ``{TTL-Based Cache Utility Maximization Using Deep Reinforcement Learning},'' in \emph{Proc. of IEEE GLOBECOM}, 2021, pp. 1--6.

\bibitem{Utility_DRL_elastic_caching}
------, ``Elastic network cache control using deep reinforcement learning,'' in \emph{Proc. of International Conference on Information and Communication Technology Convergence (ICTC)}, 2022, pp. 1006--1008.

\bibitem{Asmussen}
S.~Asmussen, \emph{Applied Probability and Queues}, ser. Stochastic Modelling and Applied Probability.\hskip 1em plus 0.5em minus 0.4em\relax Springer, 2008.

\bibitem{Berger:2014}
D.~S. Berger, P.~Gland, S.~Singla, and F.~Ciucu, ``Exact analysis of {TTL} cache networks: The case of caching policies driven by stopping times,'' in \emph{Proc. of ACM SIGMETRICS}, 2014, pp. 595--596.

\bibitem{CW_MatInv_complexity}
D.~Coppersmith and S.~Winograd, ``Matrix multiplication via arithmetic progressions,'' \emph{Journal of Symbolic Computation}, vol.~9, no.~3, pp. 251--280, 1990.

\bibitem{Walrand:Alpha_fair}
J.~Mo and J.~Walrand, ``Fair end-to-end window-based congestion control,'' \emph{IEEE/ACM Trans. Netw.}, vol.~8, no.~5, pp. 556--567, 2000.

\bibitem{byrd2000_IP_approach}
R.~H. Byrd, J.~C. Gilbert, and J.~Nocedal, ``A trust region method based on interior point techniques for nonlinear programming,'' \emph{Mathematical programming}, vol.~89, pp. 149--185, 2000.

\bibitem{IPOPT}
A.~W{\"a}chter and L.~T. Biegler, ``On the implementation of an interior-point filter line-search algorithm for large-scale nonlinear programming,'' \emph{Mathematical Programming}, vol. 106, pp. 25--57, 2006.

\bibitem{Line_search_SQP}
R.~Fletcher, S.~Leyffer, and P.~L. Toint, ``{On the Global Convergence of a Filter-SQP Algorithm},'' \emph{SIAM Journal on Optimization}, vol.~13, no.~1, pp. 44--59, 2002.

\bibitem{Line_search_IPOPT}
A.~W\"{a}chter and L.~T. Biegler, ``Line search filter methods for nonlinear programming: Motivation and global convergence,'' \emph{SIAM Journal on Optimization}, vol.~16, no.~1, p. 1–31, 2005.

\bibitem{Fricker:2012:versatile}
C.~Fricker, P.~Robert, and J.~Roberts, ``A versatile and accurate approximation for {LRU} cache performance,'' in \emph{Proc. of ITC}, 2012, pp. 1--8.

\bibitem{Gori2005}
M.~Gori, G.~Monfardini, and F.~Scarselli, ``A new model for learning in graph domains,'' in \emph{Proc. of IEEE IJCNN}, 2005.

\bibitem{Scarselli2009}
F.~Scarselli, M.~Gori, A.~C. Tsoi, M.~Hagenbuchner, and G.~Monfardini, ``The graph neural network model,'' \emph{IEEE Trans. Neural Netw.}, vol.~20, no.~1, pp. 61--80, 2009.

\bibitem{Bronstein2021}
M.~M. Bronstein, J.~Bruna, T.~Cohen, and P.~Velickovic, ``Geometric deep learning: Grids, groups, graphs, geodesics, and gauges,'' 2021, arxiv:2104.13478.

\bibitem{Li2016a}
Y.~Li, D.~Tarlow, M.~Brockschmidt, and R.~Zemel, ``Gated graph sequence neural networks,'' in \emph{Proc. of ICLR}, 2016.

\bibitem{williams1992_RL}
R.~J. Williams, ``Simple statistical gradient-following algorithms for connectionist reinforcement learning,'' \emph{Machine learning}, vol.~8, pp. 229--256, 1992.

\bibitem{Pytorch}
A.~{Paszke et al}., ``Pytorch: An imperative style, high-performance deep learning library,'' in \emph{Advances in Neural Information Processing Systems 32}.\hskip 1em plus 0.5em minus 0.4em\relax Curran Associates, Inc., 2019, pp. 8024--8035.

\bibitem{SNIA_Trace_IBM}
O.~Eytan, D.~Harnik, E.~Ofer, R.~Friedman, and R.~Kat, ``{IBM} object store traces ({SNIA IOTTA} trace set 36305),'' in \emph{SNIA IOTTA Trace Repository}.\hskip 1em plus 0.5em minus 0.4em\relax Storage Networking Industry Association, 2019.

\end{thebibliography}
